\documentclass[twocolumn]{aastex63}

\usepackage{amsmath}
\usepackage{mathtools}% http://ctan.org/pkg/mathtools
\usepackage[normalem]{ulem} % Strikethrough text \sout{}
\usepackage{natbib}
\usepackage{comment}
\usepackage{soul} 
\usepackage{amssymb}
\usepackage{xspace}
\usepackage{xifthen}

\usepackage{soul} 
\usepackage{amsmath}
\usepackage{amssymb}
\usepackage{xspace}
\usepackage{xifthen}

\newcommand{\unit}[1]{\ensuremath{\mathrm{\,#1}}\xspace}
\newcommand{\Gyr}{\unit{Gyr}}

\newcommand{\degree}{\ensuremath{{}^{\circ}}\xspace}
\newcommand{\mas}{\unit{mas}}

\newcommand{\km}{\unit{km}}
\newcommand{\kms}{\km \second^{-1}}
\newcommand{\pc}{\unit{pc}}

\newcommand{\second}{\unit{s}}

\newcommand{\Msun}{\unit{M_\odot}}

\newcommand{\e}{\unit{e^{-}}}

\newcommand{\feh}{{\rm [Fe/H]}}

\newcommand{\yr}{\unit{yr}}
\newcommand{\masyr}{\unit{\mas \yr^{-1}}}

\newcommand{\COMMENT}[3]{{}}

\newcommand{\stamp}[0]{F6\xspace}

\received{}
\revised{}
\accepted{}
\submitjournal{ApJ}

\shorttitle{Magellan/M2FS Spectroscopy of Fornax 6}
\shortauthors{A.B. Pace et al.}

\graphicspath{{./}{figures/}}

\begin{document}

\title{Spectroscopic Confirmation of the Sixth Globular Cluster in the Fornax Dwarf Spheroidal Galaxy\footnote{This paper presents data gathered with the Magellan Telescopes at Las Campanas Observatory, Chile.}}

\correspondingauthor{Andrew B. Pace}
\email{apace@andrew.cmu.edu}

\author[0000-0002-6021-8760]{Andrew B. Pace}
\affiliation{McWilliams Center for Cosmology, Carnegie Mellon University, 5000 Forbes Ave, Pittsburgh, PA 15213, USA }

\author[0000-0003-2496-1925]{Matthew G. Walker}
\affiliation{McWilliams Center for Cosmology, Carnegie Mellon University, 5000 Forbes Ave, Pittsburgh, PA 15213, USA }

\author[0000-0003-2644-135X]{Sergey E. Koposov}
\affiliation{Institute for Astronomy, University of Edinburgh, Royal Observatory, Blackford Hill, Edinburgh EH9 3HJ, UK}
\affiliation{Institute of Astronomy, University of Cambridge, Madingley Road, Cambridge CB3 0HA, UK}
\affiliation{McWilliams Center for Cosmology, Carnegie Mellon University, 5000 Forbes Ave, Pittsburgh, PA 15213, USA }

\author[0000-0003-2352-3202]{Nelson Caldwell}
\affiliation{Harvard-Smithsonian Center for Astrophysics, 60 Garden Street, MS-15, Cambridge, MA 02138, USA}

\author{Mario Mateo}
\affiliation{Department of Astronomy, University of Michigan, Ann Arbor, MI 48109, USA}

\author[0000-0002-7157-500X]{Edward W. Olszewski}
\affiliation{Steward Observatory, The University of Arizona, 933 N. Cherry Avenue, Tucson, AZ 85721, USA}

\author[0000-0002-4272-263X]{John I. Bailey, III}
\affiliation{Department of Physics, UCSB, Santa Barbara, CA 93016, USA}

\author[0000-0002-8226-6237]{Mei-Yu Wang}
\affiliation{McWilliams Center for Cosmology, Carnegie Mellon University, 5000 Forbes Ave, Pittsburgh, PA 15213, USA }

\begin{abstract}
The Fornax dwarf spheroidal galaxy has an anomalous number of globular clusters, five, for its stellar mass. 
There is a longstanding debate about  a potential sixth globular cluster (Fornax~6) that has recently been `rediscovered' in DECam imaging.  
We present new Magellan/M2FS spectroscopy of the Fornax~6 cluster and Fornax dSph.
Combined with literature data we identify $\sim15-17$ members of the Fornax~6 cluster that this overdensity is indeed a star cluster and associated with the Fornax dSph. 
The cluster is significantly more metal-rich (mean metallicity of $\overline{\rm [Fe/H]}=-0.71\pm0.05$) than the other five Fornax globular clusters ($-2.5<{\rm [Fe/H]}<-1.4$) and more metal-rich than the bulk of Fornax. 
We measure a  velocity dispersion of $5.6_{-1.6}^{+2.0}\kms$ 
corresponding to anomalously high mass-to-light of 15$<$M/L$<$258 at 90\% confidence when calculated assuming equilibrium.
Two stars inflate this dispersion and may be either Fornax field stars or as yet unresolved binary stars. Alternatively the Fornax~6 cluster may be undergoing tidal disruption.
Based on its metal-rich nature, the Fornax 6 cluster is likely younger than the other Fornax clusters, with an estimated age of $\sim2$ Gyr when compared to  stellar isochrones.
The chemodynamics and star formation history of Fornax shows imprints of major events such as infall into the Milky Way, multiple pericenter passages, star formation bursts, and/or potential mergers or interactions. 
Any of these events may have triggered the formation of the Fornax~6 cluster.
\end{abstract}


\section{Introduction} \label{sec:intro}

The Fornax dSph (hereafter, Fornax) is the fourth most massive satellite galaxy of the Milky Way, in terms of stellar mass.  It has an extended star formation history, with major star formation at old ages ($>10$ Gyr ago), a subsequent burst at $\sim4.6$ Gyr, and recent intermittent episodes  \citep{Rusakov2021MNRAS.502..642R}.
The chemodynamics of red-giant branch stars show that a relatively metal-rich stellar population is centrally concentrated and kinematically cold, and a metal-poor population is more spatially extended and dynamically hot \citep{Battaglia2006A&A...459..423B,Walker2011ApJ...742...20W}.
There is evidence for mergers in the kinematics and spatial distribution \citep{Amorisco2012ApJ...756L...2A, delPino2017MNRAS.465.3708D}.

Fornax is unusual in its large number of globular clusters (GC) and mass in GCs compared to other galaxies of similar stellar mass ($M_{\star} \sim 2\times 10^{7} \Msun$)\citep{Huang2021MNRAS.500..986H}.
In particular, the relation between number of GCs and dwarf galaxy host mass becomes stochastic for galaxies slightly larger than Fornax \citep[e.g.,][]{Forbes2018MNRAS.481.5592F}. 
There are 5 ``bright'' GCs in Fornax that have received significant study of their structural profiles \citep{Mackey2003MNRAS.340..175M}, chemistry and metallicity \citep{Strader2003AJ....125.1291S, Hendricks2016A&A...585A..86H}, and ages \citep{deBoer2016A&A...590A..35D}.
A ``sixth'' Fornax GC was first noted by \citet{Shapley1939PNAS...25..565S}; however,  subsequent studies debated its nature and  whether it was composed of stars or background galaxies  \citep{Verner1981AJ.....86..357V, Demers1994AJ....108.1648D, Stetson1998PASP..110..533S} and it fell out of the literature for some time.
Recent DECam imaging and {\it Gaia} astrometric data have shown that the Fornax~6 GC (F6) is clearly resolved into an overdensity of stars \citep{Wang2019ApJ...875L..13W}.

Besides the large number of GCs in Fornax, their survival and lack of sinking to the center of Fornax is an open question that has implications for the dark matter halo of Fornax.
If the Fornax GCs formed at their current positions they would sink to the center of Fornax in several Gyr from dynamical fiction in a cuspy dark matter halo \citep{Goerdt2006MNRAS.368.1073G}.
In contrast, if Fornax has a cored dark matter halo, the sinking time increases to $\sim10$ Gyr. 
There is other evidence from the chemodynamics and multiple stellar populations that  suggests that Fornax contains a cored dark matter halo \citep{Walker2011ApJ...742...20W, Amorisco2013MNRAS.429L..89A}.
However, the unknown formation location of the GCs can relax constraints on the underlying dark matter halo and `cuspy` solutions are not excluded \citep{Shao2020arXiv201208058S}.
The peculiar radial distribution of the GCs may be evidence for a dwarf-dwarf merger in Fornax \citep{Leung2020MNRAS.493..320L}.

Here we present Magellan/M2FS spectroscopy that confirms that the F6 globular cluster  is a coherent stellar structure, distinct from the Fornax dSph, and whose kinematics and chemistry we measure.
In Section~\ref{sec:data} we present our observations and auxiliary data and also present our velocity and metallicity measurements with the Magellan/M2FS spectroscopy.
In Section~\ref{label:results} we discuss Fornax membership, the  membership, kinematics and metallicity of F6, and discuss the nature of F6 and compare it to other GCs. In section~\ref{label:conclusion}, we conclude and summarize our results.

\section{Data}
\label{sec:data}

\subsection{M2FS Spectroscopy}

We present results from new spectroscopic observations of Fornax stars that we obtained using the Michigan/Magellan Fiber System \citep[M2FS;][]{Mateo2012SPIE.8446E..4YM} at the 6.5-m Magellan/Clay Telescope at Las Campanas Observatory, Chile.  
M2FS is a dual-channel, multi-object echelle  spectrograph that offers various options for spectral resolution and wavelength coverage, and is fed by up to 256 fibers that can be deployed over a field of diameter $0.5^{\circ}$.  We used a configuration that covered the approximate range 5130 - 5190 \AA\, at resolution $\mathcal{R}\sim 24,000$.  

\subsubsection{Target Selection}

\begin{figure}
\includegraphics[width=\columnwidth]{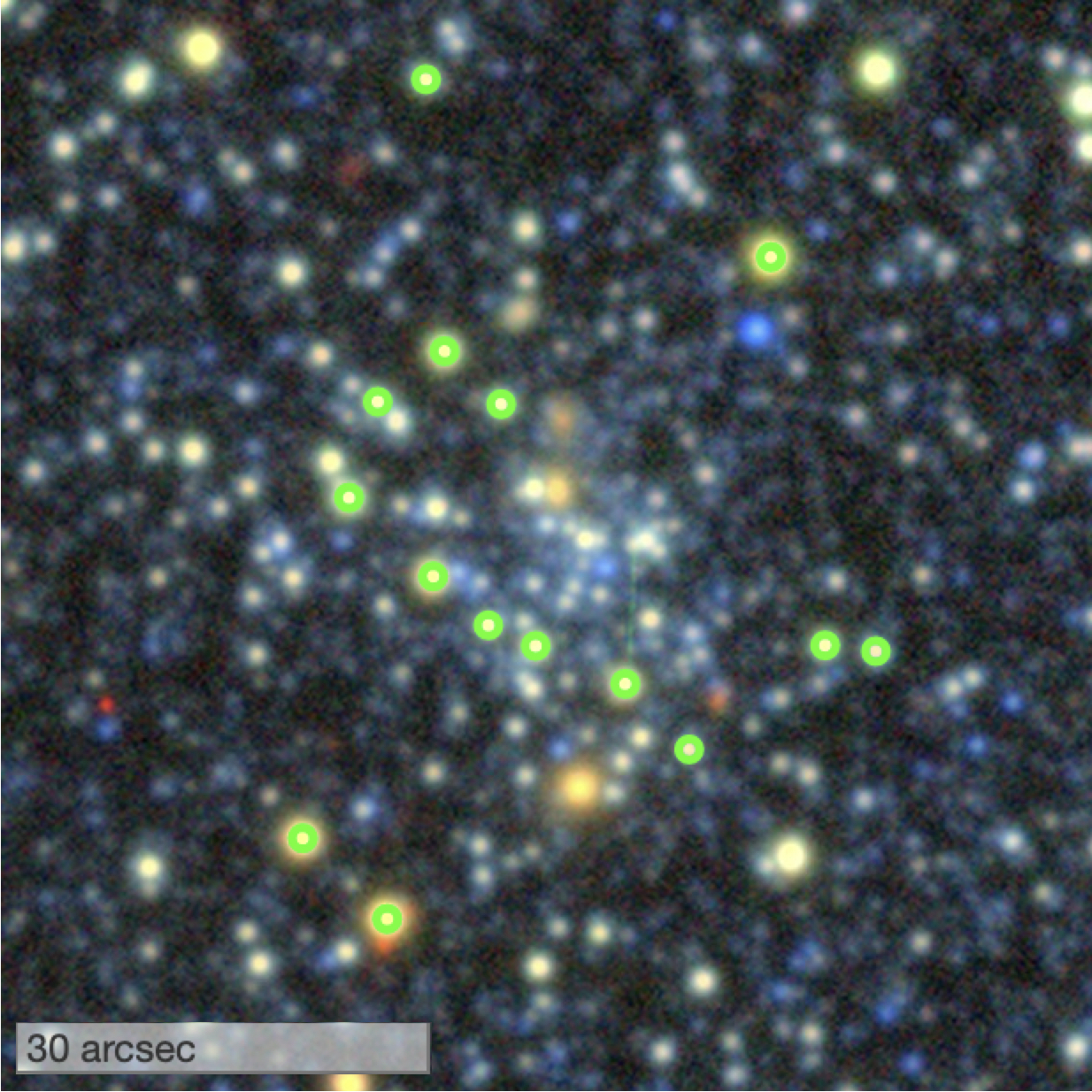}
\caption{Legacy Survey Sky Viewer false color grz coadded image of $\sim$80\arcsec by 80\arcsec  centered on the Fornax~6 cluster. The green circled stars are spectroscopic members of the Fornax~6 cluster identified in this work.}
\label{fig:cutout}
\end{figure}

We used photometry from the Dark Energy Survey (Year 1; \citealt{DES2018ApJS..239...18A_DR1}), combined with astrometry from Gaia (DR2; \citealt{GaiaBrown2018A&A...616A...1G} ) to select red giant branch (RGB) candidates along the line of sight to Fornax.  
In Figure~\ref{fig:cutout} we show a 80'' by 80'' false color grz coadded image centered on the F6 cluster made with the Legacy Survey Sky Viewer\footnote{\url{https://www.legacysurvey.org/viewer}}  \citep{Dey2019AJ....157..168D}.  The F6 members identified in this work are denoted with green circles (see Section~\ref{sec:member_f6}). 
We considered a given star to be an RGB candidate if its  $r$-band magnitude  and $g-r$ color (both corrected for extinction using the dust maps of \citealt{Schlegel1998ApJ...500..525S} with extinction coefficients from  \citealt{DES2018ApJS..239...18A_DR1})
 placed it within $\epsilon$ magnitudes of an old (age = 12 Gyr), metal-poor ([Fe/H]=-1.5) theoretical isochrone computed using the PARSEC package \citep{Bressan2012MNRAS.427..127B}.  We adopted a tolerance of $\epsilon=\sqrt{0.2^2+\sigma_{g-r}^2}$, where $\sigma_{g-r}$ is the color's observational error.  We then filtered these RGB candidates according to the parallax measured by Gaia, discarding those with nonzero parallax detected at $\ge 3\sigma$ significance.  

For the remaining RGB candidates, we assigned targeting priority based on proximity of the Gaia-observed proper motion to the Fornax mean reported by the  \citet{GaiaHelmi2018A&A...616A..12G}: $(\mu_{\alpha},\mu_{\delta})=(0.376,-0.413)$ in units of mas yr$^{-1}$.  Specifically, we awarded `points' to stars according to the following: 5 points for having proper motion within 12 km s$^{-1}$ (Fornax's line-of-sight velocity dispersion) of Fornax's mean (assuming distance 150 kpc), plus three (two, one) additional points if the proper motion is within one (three, five) times the star's proper motion uncertainty from Fornax's mean proper motion.  When allocating fibers within a given M2FS field, we randomly selected amongst the RGB candidates having the largest available point total, up to a maximum of $232$ targets, leaving up to $24$ fibers to observe regions of blank sky for the purpose of background subtraction.

\subsubsection{Observations and Data Reduction}

\begin{deluxetable*}{lll llll}
\tablewidth{0pt}
\tablecaption{ M2FS fields observed in Fornax
\label{tab:for6_obs}
}
\tablehead{$\alpha_{2000}$ (deg) & $\delta_{2000}$ (deg) & UT Date & Exp. Time (s) & $N_{obs}$ & $N_{good}$
}
\startdata
$40.04$ & $-34.59$ & 2018 Nov 26  & $3\times 1800$ & $232$ & $149$\\
$38.86$ & $-34.80$ & 2018 Nov 27  & $3\times 2000$ & $172$ & $72$\\
$39.75$ & $-34.38$ & 2018 Nov 28  & $3\times 1800$ & $232$ & $188$ \\
$39.39$ & $-35.32$ & 2018 Nov 29  & $3\times 2000$ & $232$ & $115$\\
$40.38$ & $-35.12$ & 2019 Aug 31  & $4\times 2700$ & $232$ & $104$\\
$41.06$ & $-33.86$ & 2019 Sep 2   & $2\times2400+2100+1800$ & $141$ & $60$\\
$41.06$ & $-33.86$ & 2019 Sep 3   & $4\times2400$ & $ 141$ & $41$\\
$40.32$ & $-34.59$ & 2019 Nov 24  & $2\times 3000$ & $232$ & $120$\\
$40.24$ & $-34.29$ & 2020 Jan 27  & $2\times2400+1600$ & $232$ & $123$
\enddata
% \tablecomments{ }
\end{deluxetable*}

\begin{deluxetable*}{lll cc cccc}
\tablewidth{0pt}
\tablecaption{M2FS Stellar Spectroscopy of Fornax 
\label{tab:m2fs_data}
}
\tablehead{source\_id\tablenotemark{a}&$\alpha$  & $\delta$ & HJD\tablenotemark{b} & S/N \tablenotemark{c} & $v_{\rm los}$ & [Fe/H] & $T_{\rm eff}$ & $\log_{10}{\left[g/ ({\rm cm \, s^{-2}}\right]}$ \\
&deg & deg & day & & $\kms$ & dex & K & dex 
}
\startdata
5050175928509432448 & 40.13920 & -35.01519 & 2458726.86 & 3.07 & $39.14 \pm 0.89$ & $-2.13 \pm 0.23$ & $4362.9 \pm 163.9$ & $0.66 \pm 0.33$ \\
5050176173323899136 & 40.16357 & -34.96161 & 2458726.86 & 1.89 & $52.76 \pm 0.90$ & $-1.20 \pm 0.15$ & $4060.4 \pm 50.6$ & $0.98 \pm 0.23$ \\
5050175791070483200 & 40.18383 & -35.00462 & 2458726.86 & 3.09 & $60.70 \pm 0.79$ & $-1.36 \pm 0.17$ & $4393.5 \pm 130.4$ & $0.75 \pm 0.31$ \\
5050077522218863104 & 40.19364 & -35.14245 & 2458726.86 & 1.19 & $57.77 \pm 1.78$ & $-1.26 \pm 0.65$ & $5084.1 \pm 544.0$ & $2.72 \pm 0.91$ \\
5050081817186179840 & 40.16442 & -35.04558 & 2458726.86 & 1.56 & $46.19 \pm 1.70$ & $-2.08 \pm 0.68$ & $4960.7 \pm 587.3$ & $0.93 \pm 0.67$ \\
5050078759169459968 & 40.16948 & -35.07501 & 2458726.86 & 3.05 & $55.45 \pm 0.88$ & $-1.23 \pm 0.14$ & $4208.4 \pm 118.6$ & $0.48 \pm 0.31$ \\
\enddata
\tablecomments{
\tablenotetext{a}{{\it Gaia} EDR3 source\_id}
\tablenotetext{b}{Heliocentric Julian date}
\tablenotemark{c}{Median signal-to-noise ratio per pixel}
(This table is available in its entirety in machine-readable form.) }
\end{deluxetable*}

\begin{deluxetable*}{ccc c cc cc}
\tablewidth{0pt}
\tablecaption{Combined MMFS and M2FS catalog of Fornax 
\label{tab:all_data}
}
\tablehead{\colhead{source\_id\tablenotemark{a}}& \colhead{$\alpha$}  & \colhead{$\delta$} & \colhead{N($N_{\rm M2FS}$)} & \colhead{$v_{\rm los}$} & \colhead{[Fe/H]} & \colhead{$p_{\rm dSph}$} & \colhead{$p_{\rm F6}$} \\
& deg & deg & & $\kms$ & dex & &   
}
\startdata
5062218123460339584 & 40.02271 & -34.42244 & 1(1) & $48.06 \pm 1.24$ & $-0.55\pm0.27$ & $1.00_{-0.00}^{+0.00}$ & $0.97_{-0.02}^{+0.01}$ \\
5062218127750584320 & 40.02601 & -34.42452 & 1(1) & $45.49 \pm 1.54$ & $-0.78\pm0.42$ & $1.00_{-0.00}^{+0.00}$ & $0.98_{-0.01}^{+0.01}$ \\
5062218157813975296 & 40.03333 & -34.42791 & 3(1) & $53.62 \pm 0.70$ & $-0.77\pm0.10$ & $1.00_{-0.00}^{+0.00}$ & $0.96_{-0.03}^{+0.02}$ \\
5062218157813975424 & 40.03538 & -34.42630 & 1(1) & $56.91 \pm 0.76$ & $-0.62\pm0.11$ & $1.00_{-0.00}^{+0.00}$ & $0.91_{-0.10}^{+0.04}$ \\
5062218157814038528 & 40.03194 & -34.41655 & 1(1) & $51.96 \pm 1.22$ & $-0.62\pm0.20$ & $1.00_{-0.00}^{+0.00}$ & $0.98_{-0.01}^{+0.01}$ \\
5062218157814039040 & 40.03057 & -34.41761 & 1(1) & $49.95 \pm 1.30$ & $-1.09\pm0.37$ & $1.00_{-0.00}^{+0.00}$ & $0.96_{-0.02}^{+0.01}$ \\
5062218162109182080 & 40.03356 & -34.41757 & 1(0) & $51.85 \pm 0.68$ &  & $1.00_{-0.00}^{+0.00}$ & $0.91_{-0.04}^{+0.03}$ \\
5062218162109188096\tablenotemark{b} & 40.03088 & -34.42204 & 1(1) & $36.56 \pm 1.12$ & $-0.73\pm0.25$ & $1.00_{-0.00}^{+0.00}$ & $0.96_{-0.34}^{+0.02}$ \\
5062218162115470592 & 40.03426 & -34.41948 & 2(1) & $53.47 \pm 0.94$ & $-0.56\pm0.15$ & $1.00_{-0.00}^{+0.00}$ & $0.98_{-0.01}^{+0.01}$ \\
5062218162115472000 & 40.03222 & -34.42106 & 1(1) & $52.90 \pm 0.88$ & $-0.95\pm0.12$ & $1.00_{-0.00}^{+0.00}$ & $0.96_{-0.04}^{+0.02}$ \\
5062218162115473408 & 40.02973 & -34.42245 & 1(1) & $52.95 \pm 1.10$ & $-0.47\pm0.17$ & $1.00_{-0.00}^{+0.00}$ & $1.00_{-0.00}^{+0.00}$ \\
5062218162115474560 & 40.02756 & -34.42321 & 2(1) & $51.76 \pm 0.64$ & $-0.87\pm0.16$ & $1.00_{-0.00}^{+0.00}$ & $0.99_{-0.00}^{+0.00}$ \\
5062218333907887232 & 40.02148 & -34.42256 & 1(1) & $52.42 \pm 1.62$ & $0.37\pm0.75$ & $1.00_{-0.00}^{+0.00}$ & $0.91_{-0.04}^{+0.03}$ \\
5062218368273902848\tablenotemark{c} & 40.02403 & -34.41467 & 1(1) & $41.51 \pm 0.73$ & $-0.72\pm0.10$ & $1.00_{-0.00}^{+0.00}$ & $0.92_{-0.19}^{+0.04}$ \\
5062218368275184640 & 40.03239 & -34.41112 & 1(0) & $43.85 \pm 4.06$ &  & $1.00_{-0.00}^{+0.00}$ & $0.63_{-0.12}^{+0.09}$ \\
\enddata
\tablecomments{\tablenotetext{a}{{\it Gaia} EDR3 source\_id}
\tablenotetext{b}{star1}
\tablenotetext{c}{star2}
(This table is available in its entirety in machine-readable form.)  }
\end{deluxetable*}

Between the years 2018 and 2020, we observed 8 Fornax fields with M2FS.  Table \ref{tab:for6_obs} lists central coordinates (which coincide with the location of a bright star used for wavefront sensing) for each field, along with the date of observation, total exposure time, number of targeted stars and number of stars with successful measurements.  
Four fields targeted the  F6 cluster and surrounding Fornax dSph and four targeted the outskirts of the galaxy.

We process all M2FS spectra using a custom-built  Python-based pipeline that, where applicable, incorporates modules that are publicly available as part of the  Astropy software package \citep{Astropy2013A&A...558A..33A, Astropy2018AJ....156..123A}.  Complete details will be provided by Walker et al. (in preparation).  Briefly, we use standard procedures to perform overscan, bias, dark current and gain corrections.  We identify and trace the 2D aperture corresponding to each spectrum, and then extract the  one-dimensional spectra by taking the weighted averages of pixel counts along the direction approximately orthogonal to the dispersion axis.  We use spectra acquired during twilight to perform flat-field and wavelength-dependent throughput corrections, and we use spectra of a Thorium-Argon-Neon arc lamp, acquired before and after each science exposure, to perform wavelength calibration.  We propagate the variance in each pixel through all processing steps.  

Following the procedure of \citet{Walker2015MNRAS.448.2717W,Walker2015ApJ...808..108W,
Walker2016ApJ...819...53W}, we model each spectrum using a library of synthetic templates that cover a grid with dimensions of effective temperature $T_{\rm eff}$, surface gravity $\log g$, and [Fe/H] metallicity.  In addition to these stellar-atmospheric parameters, we also obtain estimates of the line-of-sight velocity, $v_{\rm los}$.  After imposing a quality control filter that requires  median signal-to-noise ratio S/N$>0$ (sky over-subtraction can result in unphysically negative S/N) and velocity error $\epsilon_v<5$ km s$^{-1}$, the M2FS sample includes 980 observations of 804 unique Fornax stars.   
The resulting M2FS catalog for Fornax is listed in Table~\ref{tab:m2fs_data}.

\subsection{Systematic errors}
In order to determine systematic errors for the velocities and metallicities determined from M2FS spectra, we use repeat measurements of all stars observed with M2FS (using the same instrument configuration as for Fornax); Walker et al. (in preparation) will provide the complete catalog resulting from these observations.    The targets are primarily dSphs but also include several globular clusters. 
Overall, there are 14437 pairs and 5646 pairs where both measurement passes our quality cuts. 
We model the pair-wise velocity differences  as a mixture of a Gaussian with an outlier model \citep[Section 4.1 of][]{Li2019MNRAS.490.3508L}.
The final uncertainty is treated as a systematic error plus a scaling parameter, $\sigma_{v, calib}^2 = \sigma_{v, \, {\rm systematic}}^2 + (k_v \sigma_{v, {\rm obs}})^2$, where $\sigma_{v, {\rm obs}}$ is taken from the previous section.
For the velocity and metallicity systematic errors, we find $k_v=0.95 \pm 0.02$, $\sigma_{v, \, {\rm systematic}}=0.59\pm0.02 \kms$,  $k_{\rm [Fe/H]}=1.08\pm0.03$, and $\sigma_{{\rm [Fe/H]}, \, {\rm systematic}}=0.04\pm0.01$. 
We adopt the median values for our analysis.

We have compared the M2FS [Fe/H] metallicity to several other large spectroscopic surveys  \citep{Battaglia2006A&A...459..423B, Kirby2010ApJS..191..352K, Letarte2010A&A...523A..17L, Lemasle2012A&A...538A.100L, Hill2019A&A...626A..15H, Hendricks2014A&A...572A..82H, Theler2020A&A...642A.176T}.  This comparison  includes M2FS spectroscopy for the Carina, Sculptor, and Sextans dSphs which will be presented in future work  (Walker et al. in preparation). 
We found overlaps of 40-200 stars between the M2FS sample and literature data. In all cases, the M2FS [Fe/H] was  more metal-poor than the literature sample by $\Delta {\rm [Fe/H} \sim -0.1- -0.3~{\rm dex}$. 
We used the sample with the  largest overlap \citep{Kirby2010ApJS..191..352K} to set the [Fe/H] zero-point  and we  find $\Delta {\rm [Fe/H} \sim -0.2~{\rm dex}$.
Our previous work with the same spectral modeling methodology  determined the [Fe/H] zero-point by fitting twilight spectra \citep{Walker2015MNRAS.448.2717W,Walker2015ApJ...808..108W,
Walker2016ApJ...819...53W}. The offset from twilight spectra is $\Delta {\rm [Fe/H} = -0.32~{\rm dex}$  which is similar in magnitude and sign to the offset we find comparing to other spectroscopic metallicity measurements.
Both the systematic errors and [Fe/H] zeropoint are included in Table~\ref{tab:m2fs_data}.

\subsection{Additional Data}

We use photometric and astrometric data from the {\it Gaia} EDR3 catalog \citep{Gaia_Brown_2021A&A...649A...1G}.
We only utilize astrometric data that passed the following cuts: {\tt ruwe} $<1.4$ \citep{Gaia_Lindegren_2021A&A...649A...2L} and $|C^*| \le 3 \sigma_{C^*}(G)$ \citep{Gaia_Riello2021A&A...649A...3R}. 
We include additional   Magellan/MMFS spectroscopic data from \citet{Walker2009AJ....137.3100W} which increases the number of Fornax members by $\sim2400$ stars.  This data is ideal improving the Fornax dSph velocity distribution near F6 to assist with the separation between the dSph and F6 stars.

\subsection{Final Spectroscopic Sample}

In order to combine the MMFS and M2FS samples we must correct for zero point velocity shifts and verify that the velocity errors of the two samples are not biased relative to one another.  Finding a small velocity offset between two different catalogs is not uncommon, given different instruments and methodology for velocity measurements.
Furthermore, previous analysis has concluded that the errors in the Fornax MMFS data set are underestimated \citep{Minor2013ApJ...779..116M}.  There are 4113 velocity measurements of 3304 stars with the combined MMFS and M2FS samples and 
there are 200 pairs of measurements between MMFS and M2FS.
We compare repeat velocities between MMFS and M2FS using the normalized difference between two observations, $\Delta=(v_1-v_2)/\sqrt{\sigma_{v,1}^2+\sigma_{v,2}^2}$, and find that $\overline{\Delta_v} = -0.16 \kms$ and $\sigma_{\Delta_v} = 1.2 \kms$ between the repeat measurements. 
After applying an offset of $-0.4545 \kms$ and scaling the errors by a factor of $1.355$ to the MMFS data we find $\overline{\Delta_v} = 0$ and $\sigma_{\Delta_v} = 1 \kms$.  
Out increase in error of the MMFS Fornax velocities by a factor of 1.355 is similar to the correction applied by \citet{Minor2013ApJ...779..116M}.

To determine the systemic velocity and error of stars with  multiple measurements, we fit the multi-epoch data with a normal distribution with a free mean and dispersion parameter. 
The mean parameter and statistics on its posterior distribution are used for combined  velocity  measurement.
We use the median and the 84, 16\% confidence intervals for the measurement and error, respectively.
This process will account for stars with evidence of velocity variation as the mean will have a larger error than the individual measurements (representing our ignorance of where it has been observed in its period, in the case of binaries).
We have compared this to  other common methods for combined velocity measurements, such as computing the weighted mean velocity, and either the variance of the weighted mean or standard deviation of the mean velocity for the measurement error.  The variance of the weighted mean ($\left(\Sigma_i \,\sigma_{v,i}^{-2}\right)^{-1/2}$) underestimates the measurement error if there is velocity variability.  
In this manner our combined measurements are more robust to velocity variation.

Many MW stars can be immediately identified with {\it Gaia} EDR3 astrometry \citep[e.g.,][]{Pace2019ApJ...875...77P}.
In particular, stars with non-zero parallax, $\varpi - 4\sigma_{\varpi}>0$, and large proper motion (relative to a star at the Fornax dSph distance) are nearby dwarf stars in the Galactic foreground.
We identify stars with large velocities by comparing the tangential velocity to the MW escape velocity assuming the star is at the distance of Fornax.
The escape velocity computed with \texttt{galpy} using the potential \texttt{MWPotential2014}
with a larger halo mass, $M_{\rm vir}=1.6\times 10^{12} \Msun$ \citep{Bovy2015ApJS..216...29B}.
We then compute $v_{\rm tan}$ from the proper motion, assuming all stars are at the distance of Fornax and  account for the Sun’s reflex motion, assuming $({\rm U_\odot,\, V_\odot, \, W_\odot}) = (11.1, 12.24, 7.25)\kms$, with a circular velocity of $220\kms$ \citep{Schonrich2010MNRAS.403.1829S}.
We consider stars to be MW foreground stars with a loose cut of $v_{\rm tan} - 4.5 \sigma_{v_{\rm tan}}<v_{\rm escape}$.
MW stars identified in this manner are not excluded from the modeling but are fixed as MW stars to help determine the MW velocity and metallicity distribution \citep[e.g.,][]{Pace2020MNRAS.495.3022P}.
Overall, there are 167 MW stars identified in this manner.
We further apply a loose color-magnitude selection in $G-G_{RP}$ vs $G$ based on candidate members selected by velocity and proper motion. 
The combined M2FS and MMFS catalog is included in Table~\ref{tab:all_data}.  

\section{Results and Discussion}
\label{label:results}

\begin{table*}[]
    \centering
\begin{tabular}{l c c r}
\hline
Parameter & Fornax dSph & Fornax 6  & citations \\
\hline
R.A. (J2000, deg) & 39.9583	& 40.02875 & W19a, W19b \\
Dec (J2000, deg) & -34.4997 & -34.422 & W19a, W19b \\
$\epsilon$ & $0.31\pm0.002$ &$0.41\pm 0.10$ & W19a, W19b\\ 
$\theta$ (deg) &  $42.2\pm0.2$ &$13.1_{-7.3}^{+10.4}$  & W19a, W19b\\ 
$r_h$ (arcmin/arcsec) & $19.9\pm0.06$\footnote{arcmin} & $16.8 \pm 2.0$\footnote{arcsec} & W19a, W19b \\ 
$r_h$ (parsec) &$852\pm3$ & $11.3 \pm 1.4$ & W19a, W19b\\ 
$r_c$ (arcmin) & $20.3\pm0.1$ & - & W19a\\ 
$r_t$ (arcmin) & $77.5\pm0.4$ & - & W19a\\ 
$M_V$ (mag) & $-13.46\pm0.14$ & $-4.8 \pm 0.4$ & M18, W19\\ 
\hline
$\overline{v_{\rm los}} \, (\kms)$ & $54.7_{-0.2}^{+0.2}$ & $50.5_{-1.7}^{+1.7}$ & This Work \\
$\sigma_v \, (\kms)$ & $12.1_{-0.2}^{+0.2}$ & $5.6_{-1.6}^{+2.0}$ & This Work \\
$\overline{{\rm [Fe/H]}}$ (dex) & $ -1.22_{-0.02}^{+0.02}$ &  $-0.71_{-0.05}^{+0.05}$ & This Work \\
$\sigma_{\rm [Fe/H]}$ (dex) & $0.48_{-0.02}^{+0.02}$ & $0.03_{-0.02}^{+0.06}$\footnote{$\sigma_{\rm [Fe/H]}<0.17$ at 95\% confidence level.} & This Work \\
N  & $2994.9\pm0.9$ & $17.3\pm2.7$ & This Work \\
$\overline{\mu_{\alpha \star}}$ $(\masyr)$  & $0.382_{-0.002}^{+0.002}$ & $0.392\pm0.026$  & This Work \\
$\overline{\mu_{\delta }}$ $(\masyr)$ & $-0.362_{-0.003}^{+0.003}$ & $-0.448\pm0.042$ & This Work \\
\hline
 \end{tabular}
    \caption{Literature and derived properties of the Fornax dSph and the Fornax~6 Globular cluster. 
    The number of members (N) for the dSph includes both dSph and cluster members.
    N=2989 with $p_{\rm dSph}>0.9$ and  N=14,15 with $p_{\rm F6} > 0.5, 0.9$.
    The citations are: W19a \citep{Wang2019ApJ...881..118W},
    W19b \citep{Wang2019ApJ...875L..13W}, 
    M18 \citep{Munoz2018ApJ...860...66M}.
    $r_h$ is the half-light radius along the major axis.
    }
    \label{tab:for6_properties}
\end{table*}

\begin{figure}
\includegraphics[width=\columnwidth]{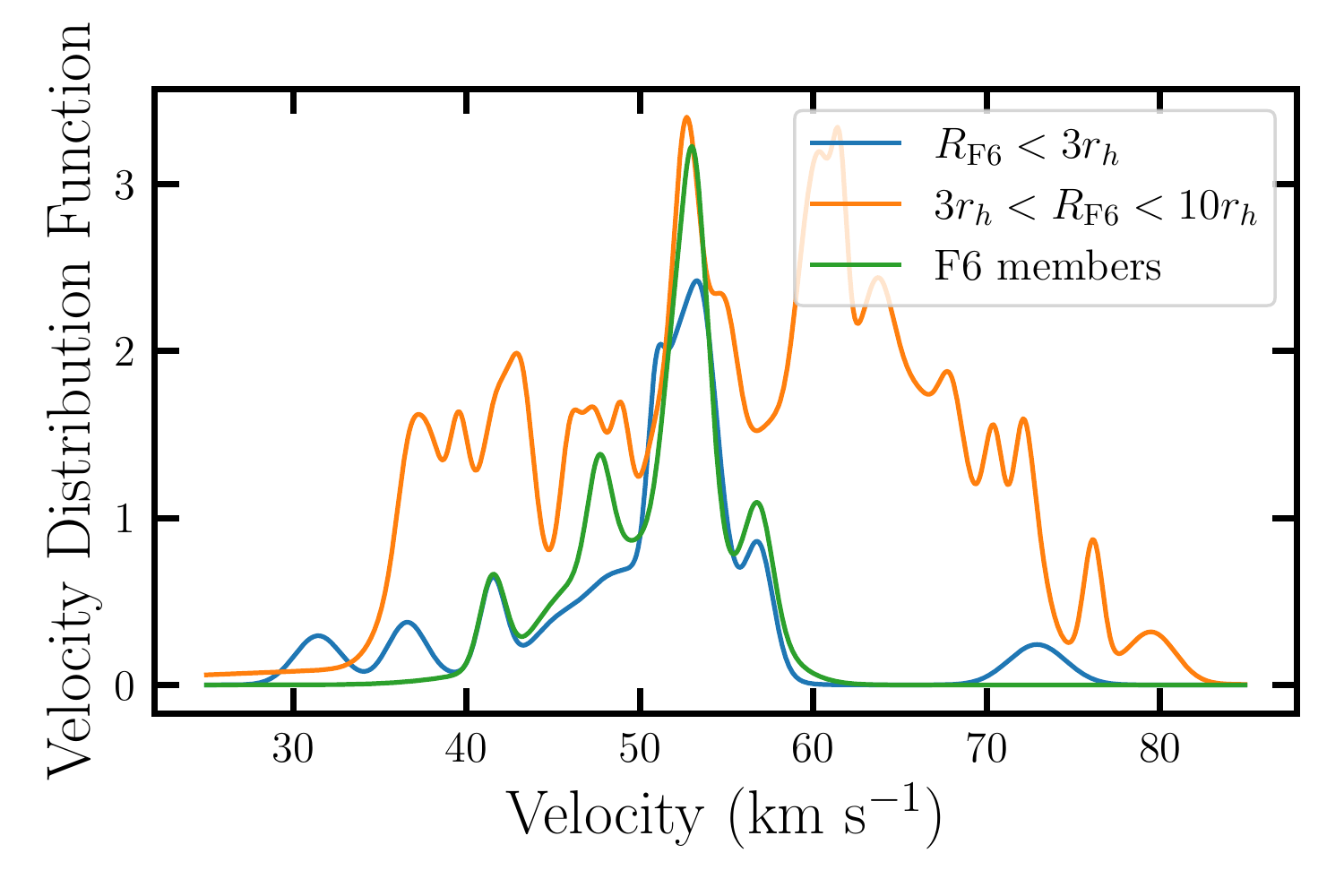}
\caption{Velocity distribution of stars near Fornax~6 ($R_{\rm F6} < 10\times r_{h, {\rm F6}}$). 
The distributions are created by smoothing each star's velocity by its measurement error.
Fornax dSph stars are shown in orange ($3\times r_{h, {\rm F6}} < R_{\rm F6} < 10\times r_{ h, {\rm F6}}$), candidate Fornax~6 members are in blue line ($R_{\rm F6} < 3\times r_{h, {\rm F6}}$), and F6 members according to our mixture model are shown in green. }
\label{fig:vel_hist}
\end{figure}

With the combined MMFS and M2FS spectroscopic data sets, there are 104 stars separated from the center of F6 by $R_{\rm F6}<10\times r_{h, \, {\rm F6}}$ of which 7 are likely MW foreground stars\footnote{We will use the subscript F6 to refer to properties of the cluster (e.g., $r_{h, \, {\rm F6}}$ is the half-light radius of F6) or measurements relative to the cluster (e.g., $R_{\rm F6}$ is the projected radial distance from the center of F6).}.
The velocity distribution of stars within $R_{\rm F6}<3\times r_{h, \, {\rm F6}}$ is more tightly clustered than the stars with $R_{\rm F6}>3\times r_{h, \, {\rm F6}}$ as shown in Figure~\ref{fig:vel_hist}.
The nearby stars cluster around $v_{\rm los}\sim 51 \kms$ which is $\sim 4 \kms$ away from the Fornax dSph mean velocity.

To be more quantitative we explore two mixture models to identify foreground MW, Fornax dSph, and F6 stars.
The first is a mixture model between the MW and the Fornax system (Section~\ref{sec:dsph_membership}).
It is clear from Figure~\ref{fig:vel_hist}, that the F6  radial velocity does not have a large offset relative to the Fornax dSph.
The small number of F6 stars in the sample will not influence the properties of the dSph (there are only 23 stars within $R_{\rm F6}<3 \times r_{h, {\rm F6}}$).
The second mixture model is then applied to the Fornax system members and separates the F6 stars from the dSph (Section~\ref{sec:member_f6}).

\subsection{Fornax System Membership}
\label{sec:dsph_membership}

\begin{figure*}
\includegraphics[width=\textwidth]{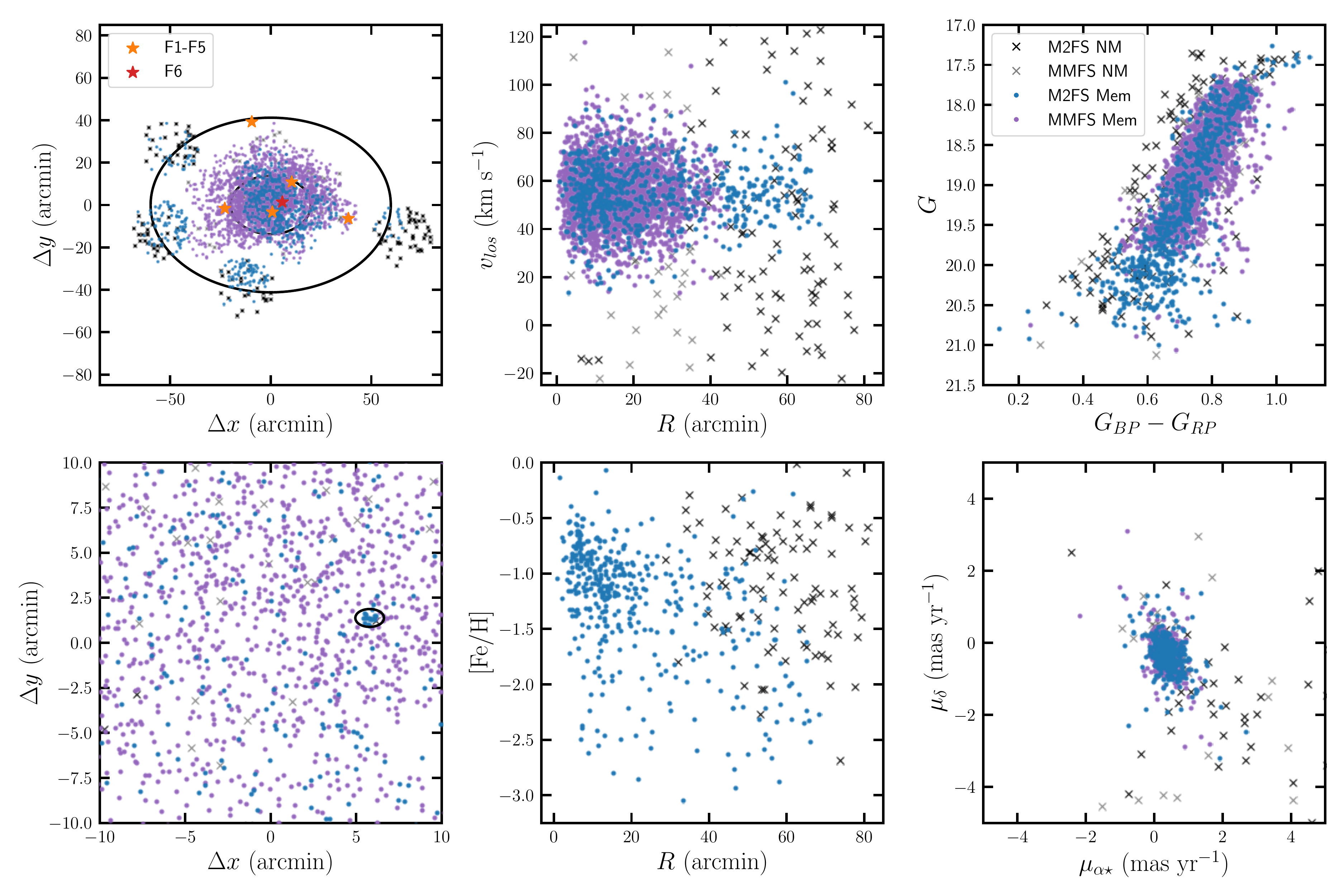}
\caption{Summary of the Magellan/M2FS and Magellan/MMFS Fornax sample. 
The points are: M2FS members (blue), MMFS members (purple), M2FS non-member (black x), and MMFS non-member (grey x). 
Top panels: (left) the spatial distribution of the entire sample. The coordinate system is rotated such that the major axis aligns with the x-axis.  Two ellipses at $r_h$  and $3\times r_h$ are shown. The Fornax globular cluster system is overlaid, orange (F1-F5) and red (F6).  (center) projected radial distance versus radial velocity (km/s), (right) $ G_{BP} - G_{RP}, G $ color-magnitude diagram.
Bottom panels: 
(left) radial velocity (km/s) versus metallicity for the M2FS sample. 
 (middle) projected radial distance versus metallicity for the M2FS sample.  
(left) {\it Gaia} EDR3 proper motions ($\mu_{\alpha \star}$ vs $\mu_\delta$). 
}
\label{fig:fornax_members}
\end{figure*}

We use a conditional likelihood based mixture model  \citep[e.g.,][]{Martinez2011ApJ...738...55M, Horigome2020MNRAS.499.3320H}, 

\begin{equation}
\mathcal{L}( \alpha| R)  = f(R) \mathcal{L}_{{\rm dSph}}(\alpha|R) + (1-f (R)) \mathcal{L}_{\rm MW} (\alpha|R) \\ 
\end{equation}

\noindent where $\alpha$ is the data vector ($\alpha$ = $\left\{v_{\rm los}\right. $, $\mu_{\alpha \star}$, 
$\mu_{\delta}$, ${\rm [Fe/H] }$ $\left. \right\}$) and $f(R)$ is the local membership probability and is given by: 
\begin{equation}
f(R) = \Sigma_{\rm dSph}(R)/(\Sigma_{\rm dSph}(R) + \Sigma_{\rm MW}(R)). \\ 
\end{equation}

\noindent Here $\Sigma (R)$ is the projected stellar density distribution.
We model the conditional likelihood instead of the full likelihood due to the difficulty in modeling the  spectroscopic selection function.

For both components, we model the velocity and metallicity distributions as normal distributions and  the proper motion distribution as multi-variate normal distribution.
For the dSph proper motion distribution we fix the dispersion to $12 \kms$ as {\it Gaia} EDR3 proper motions are not precise enough to measure internal motions ($\sim0.017 \masyr$).
For the dSph spatial distribution, we model it with a King radial profile \citep{King1962AJ.....67..471K}:

 \begin{equation}
	\Sigma_{\star} (R_e) \propto \left[ \left(1 + \frac{R_e^2}{r_c^2} \right)^{-1/2} - \left(1 + \frac{r_t^2}{r_c^2} \right)^{-1/2} \: \right]^2
 \end{equation}

\noindent Where $R_e^2=x^2 + y^2/(1-\epsilon)^2$ is the elliptical radius, $r_c$ is the core radius, $r_t$ is the tidal radius, and $\epsilon=1-b/a$ is the ellipticity. Here $x$ and $y$ are along the major and minor axis respectively, which are rotated with respect to the on-sky plane by the position angle, $\theta$. 
For the spatial parameters,  we assume the following Gaussian priors based on deep photometric measurements from \citet{Wang2019ApJ...881..118W}: $\epsilon=0.31\pm0.002$, $\theta=42.2\pm0.2 \deg$, $r_c=22.3\pm0.1 \arcmin$, and $r_t=77.5\pm0.4 \arcmin$.
For brighter dSphs, two parameter radial density profiles are found to be better fits \citep{Munoz2018ApJ...860...66M} and the outer regions of the Fornax dSph are found to fall off more steeply than a Plummer radial distribution \citep{Moskowitz2020ApJ...892...27M}.
We assume the MW foreground stars are uniformly distributed within the Fornax region.

Overall, the 10 free parameters for the dSph distribution are: $\overline{v}$ and $\sigma_v$ (velocity), $\overline{\rm [Fe/H]}$ and  $\sigma_{\rm [Fe/H]}$ (metallicity) and $\overline{\mu_{\alpha \star}}$, $\overline{\mu_{\delta}}$ (proper motion), and $\epsilon$, $\theta$, $r_c$, and $r_t$ (spatial). 
The dispersion terms ($\sigma_v$ and $\sigma_{\rm [Fe/H]}$) are modeled with Jeffreys (log-uniform) priors, the mean parameters with uniform priors and the spatial parameters with Gaussian priors. 
The 8 parameters for the MW model are:  $\overline{v_{MW}}$ and $\sigma_{v, MW}$ (velocity), $\overline{\rm [Fe/H]_{MW}}$, $\sigma_{\rm [Fe/H], MW}$, $\overline{\mu_{\alpha \star, MW}}$, $\overline{\mu_{\delta, MW}}$,$\sigma_{\mu_{\alpha \star, MW}}$,  $\sigma_{\mu_{\delta, MW}}$.  
Similar to the dSph parameters, the dispersion parameters are modeled with Jeffreys priors and the mean parameters with uniform priors.
There is one additional parameter, $N_{dSph}/N_{MW}$, which is relative normalization parameter in the fraction term.
To determine membership, $p$, we compute the ratio of the dSph likelihood to total likelihood  \citep{Pace2020MNRAS.495.3022P}.

The properties we find for the  Fornax dSph are:
$\overline{v} = 54.7_{-0.2}^{+0.2} \kms$,
$\sigma_v = 12.1_{-0.2}^{+0.2} \kms$,
$\overline{\rm [Fe/H]} = -1.22_{-0.02}^{+0.02}$,
$\sigma_{\rm [Fe/H]} = 0.48_{-0.02}^{+0.02}$,
$\overline{\mu_{\alpha \star}} = 0.382_{-0.002}^{+0.002} \masyr$, and
$\overline{\mu_{\delta}} = -0.362_{-0.003}^{+0.003} \masyr$.
In total we find 2989 stars with a membership probability $p_{\rm dSph}>0.9$ which we consider our membership threshold.
We include  the membership probabilities ($p_{\rm dSph}$)  in Table~\ref{tab:all_data}.
In Figure~\ref{fig:fornax_members}, we show the spatial, velocity, metallicity, and proper motion distributions of our Fornax sample.  

The values for $\overline{v}$ and $\sigma_v$ are similar to previous results with the MMFS sample \citep{Walker2009ApJ...704.1274W} and other Fornax kinematic studies \citep[e.g.,][]{Battaglia2006A&A...459..423B}.
The new M2FS observations increase the number of members known at large radius. 
We identify 165 members between $2 < R/r_h < 3.4$.
We find Fornax to be more metal-poor than previous Keck/DEIMOS data by $\Delta [Fe/H] \sim -0.2$ \citep{Kirby2013ApJ...779..102K}. The classical satellites are known to contain metallicity gradients \citep{Kirby2011ApJ...727...78K} and the mean metallicity difference is likely due to our more extended sample, tracing the outer regions. 
The median elliptical radius is $R_{\rm ell}/r_h\sim0.4$ and $R_{\rm ell}/r_h\sim1$ for the Keck/DEIMOS and Magellan/M2FS data sets, respectively.
When we add a metallicity gradient option to the model (${\rm [Fe/H]}(R)={\rm [Fe/H]}_0 + m R$), we find that the radial gradient is quite large, $m=-0.28\pm0.02 \,{\rm dex \, R_h^{-1}}$ or $m=-0.84\pm0.07 \, {\rm dex \, deg^{-1}}$.
This is  larger than the (unresolved) value reported by  \citet{Kirby2011ApJ...727...78K} and is again due to our extended spatial sample.

The radius versus metallicity and velocity versus metallicity panel of Figure~\ref{fig:fornax_members} show evidence for chemodynamical sub-populations in Fornax. The metal-rich component is more centrally concentrated and kinematically colder whereas the metal-poor component is more extended and kinematically hot.  
This has been previously observed in Fornax \cite[e.g.][]{Battaglia2006A&A...459..423B, Walker2011ApJ...742...20W} and will be explored further in furture works.

Our proper motion measurements agrees with other {\it Gaia} EDR3  measurements \citep{McConnachie2020RNAAS...4..229M, Vitral2021MNRAS.504.1355V}. 
Our results differ within the statistical errors to previous {\it Gaia} DR2 measurements \citep{GaiaHelmi2018A&A...616A..12G, Fritz2018A&A...619A.103F}, with the offset likely due to proper motion zero-point systematic errors in the DR2 catalog. 
These systematic errors have decreased by at least a factor of two in the EDR3 catalog.

\subsection{Membership and Properties of Fornax 6}
\label{sec:member_f6}

\begin{figure}
\includegraphics[width=\columnwidth]{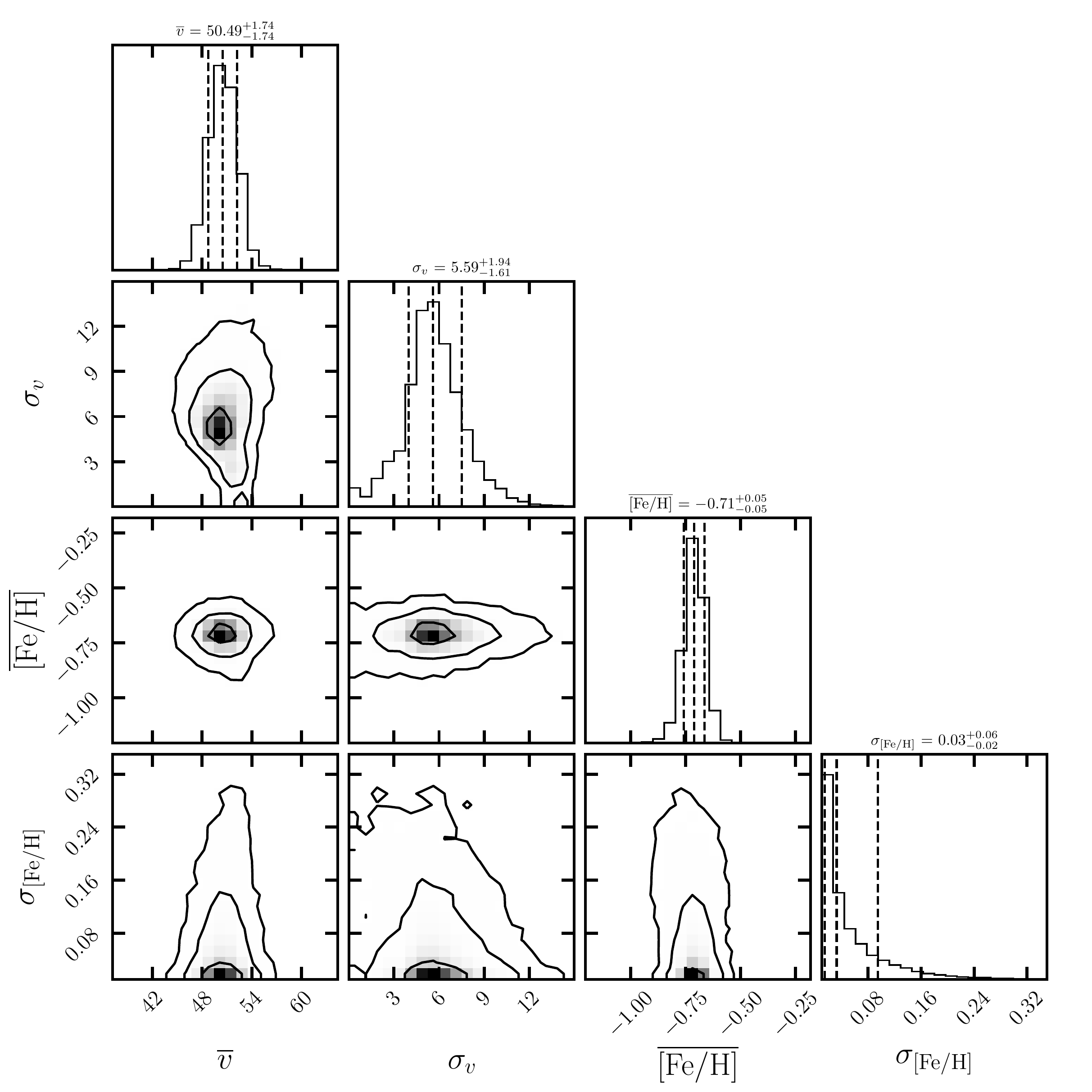}
\caption{Corner plot of properties of Fornax 6 based on the model 1 mixture model.  The parameters from left to right are:  mean velocity ($\overline{v}$, $\kms$),  velocity dispersion ($\sigma_{v}$, $\kms$), mean metallicity ($\overline{\rm [Fe/H]}$), and metallicity dispersion ($\sigma_{\rm [Fe/H]}$).
Both dispersions are modeled with Jeffreys priors and the metallicity dispersion is constrained to be $\sigma_{\rm [Fe/H]}<0.16$ at 95\% confidence interval.
}
\label{fig:for6_corner}
\end{figure}

\begin{figure*}
\includegraphics[width=\textwidth]{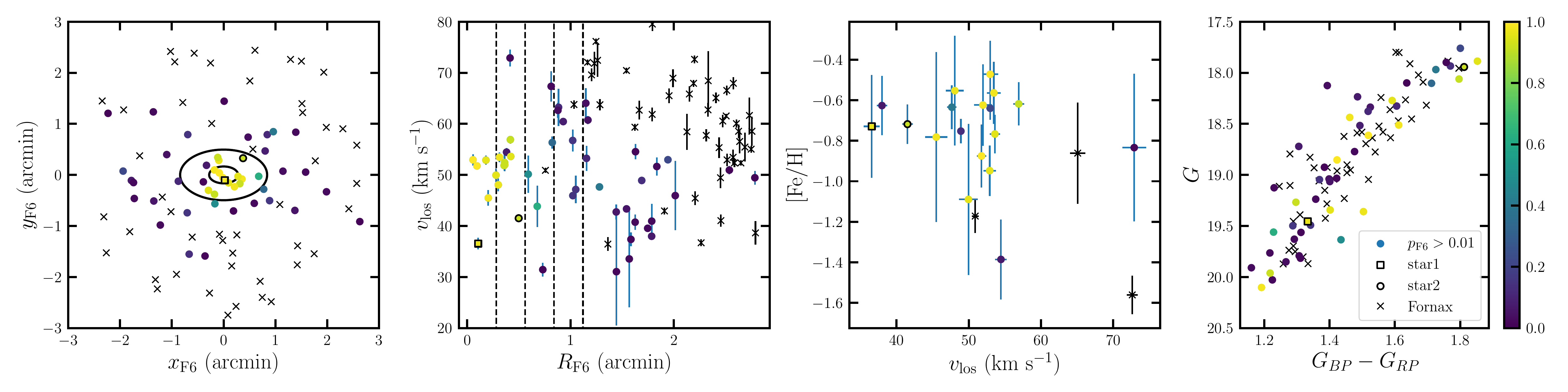}
\caption{Properties of Fornax~6 members compared to nearby Fornax dSph members (only stars with $r_{\rm F6} < 10\times r_{h, {\rm F6}}$ are included). The color bar shows stars with membership probabilities between $0.01< p_{\rm F6} < 1$ and stars with smaller membership probability are Fornax dSph and shown with an x.
(left) The spatial distribution of stars relative to the center of F6 (in arcmin).  The x-axis is aligned with the F6 major axis and ellipses are shown at $r_{ h, {\rm F6}}$ and $3 \times r_{ h, {\rm F6}}$.
(middle-left) Radial distance from F6 center (arcmin)  versus radial velocity (km/s).  The vertical lines are multiples of 1 to 4 of  $r_{ h, {\rm F6}}$.
(middle-right) Radial velocity (km/s) versus metallicity.
(right) $G_{BP}-G_{RP}$ vs $G$ color-magnitude diagram ({\it Gaia} EDR3 photometry).
Two stars with more questionable membership are identified with a black square (star1) and circle (star2).  Both are outliers in radial velocity compared to the other F6 members.
}
\label{fig:for6_diagnostic}
\end{figure*}

To determine the properties and membership of F6, we utilize a  mixture model composed of the Fornax dSph and F6  with all stars identified as Fornax members in the previous model ($p_{\rm dSph}>0.9$). 
We do not include  proper motion measurements  as {\it Gaia} EDR3 measurements are not precise enough to differentiate between the cluster and dSph due to the large distance of the system.
The F6 spatial distribution is modeled as a Plummer distribution \citep{Plummer1911MNRAS..71..460P}. Similar to the dSph model, the spatial parameters are varied with Gaussian priors based on the parameters from \citet{Wang2019ApJ...875L..13W} and are listed in Table~\ref{tab:for6_properties}.
The velocity and metallicity distributions of F6 are modeled with Gaussian distributions.

The properties of the F6 cluster we find are:  
$\overline{v}_{\rm F6} = 50.5_{-1.7}^{+1.7} \kms$,
$\sigma_{v, {\rm F6}} = 5.6_{-1.6}^{+2.0} \kms$,
$\overline{\rm [Fe/H]}_{\rm F6} = -0.71_{-0.05}^{+0.05}$, and
$\sigma_{\rm [Fe/H], {\rm F6}} = 0.03_{-0.02}^{+0.06}$.
The properties of the Fornax dSph are unchanged compared to the previous model.
The corner plot of these parameters is shown in Figure~\ref{fig:for6_corner}.
We find a total membership of $\Sigma p_{\rm F6} = 17.3\pm2.7$ in the cluster, with 15 and 14 stars having membership larger than 0.5 and 0.9, respectively.
The F6 parameters are summarized in Table~\ref{tab:for6_properties}.
From the members we identify, we measure the systemic proper motion of F6 to be  $\overline{\mu_{\alpha \star} }_{\rm F6}=0.392\pm0.026 \masyr$ and 
$\overline{\mu_{\delta }}_{\rm F6}=-0.448\pm0.042 \masyr$ .
The F6 systemic proper motion is consistent with the Fornax dSph after considering the large errors. We note that the Fornax dSph systemic proper motions errors are $\sim 1\kms$ whereas the F6 errors are  $\sim25\kms$.

In Figure~\ref{fig:for6_diagnostic}, we summarize the properties of stars near F6 ($R_{\rm F6}<10\times r_{ h, {\rm F6}}$) and show their spatial distribution, velocity distribution, metallicity distribution, and location on color-magnitude diagrams colored by their membership identified in the model.
The F6 members are tightly clustered within $R_{\rm F6}<3\times r_{ h,{\rm F6}}$ and around ${\rm [Fe/H]}\sim -0.7$ and $v \sim 50 \kms$. 
We have included the F6 members ($p_{F\rm 6}>0.5$) in the false color grz image centered on F6 in Figure~\ref{fig:cutout}.
The distribution of Fornax~6 members is similar to the Fornax dSph in the color-magnitude diagram( $G_{BP}-G_{RP}, G$).
There are two stars within $2\times r_{h,{\rm F6}}$ considered F6 non-members.  
The first is considered a non-member due to its large velocity offset to F6 ($v\sim82 \kms$) and the second is due to its  metallicity relative to the F6 mean ($[Fe/H]\sim-1.2$).

The velocity distribution of F6 (Figure~\ref{fig:vel_hist}) is non-Gaussian.  The majority of the stars are peaked at $v\sim52\kms$ and there are $\sim5$ stars with $32 \lesssim v \lesssim 50 \kms$.  All stars with smaller velocity that contain metallicity measurements match the metallicty of the cluster.
We explore the two most extreme outliers in more detail.
We will refer to these two stars as star1 ({\it Gaia} EDR3 source\_id=5062218162109188096, 
$R_{for6}/r_{ h, {\rm F6}}=0.4$, $v=36.6\pm1.1$, $\feh=-0.73\pm0.25$) and star2 ({\it Gaia} EDR3 source\_id=5062218368273902848, $R_{\rm F6}/r_{ h, {\rm F6}}=1.8$, $v=41.5\pm0.7$, $\feh=-0.72\pm0.10$).
Both are close to the cluster and have a metallicity that matches with the cluster but are $2.6\sigma$ and $1.5\sigma$ outliers in velocity. 
If we exclude star1 and recompute the kinematics we find $\overline{v}_{\rm F6} = 50.9_{-1.2}^{+1.1} \kms$ and
$\sigma_{v, {\rm F6}} = 4.0_{-0.8}^{+1.1} \kms$ and if we exclude both we find $\overline{v}_{\rm F6} = 51.7_{-0.9}^{+0.8} \kms$,
$\sigma_{v, {\rm F6}} = 2.7_{-0.6}^{+0.9} \kms$.
The exclusion of these two stars then changes the  kinematics by $\Delta \overline{v}_{\rm F6}\sim 0.3-2.0 \kms$ and $\Delta \sigma_{v, {\rm F6}} \sim 1.7-3.0 \kms$. 
When star1 is excluded, it is a $3.2\sigma$ outlier ($4.4\sigma$ when both star1 and star2 are excluded) from Fornax~6 and sigma-clipping algorithms typically exclude stars at greater than $2.58-3\sigma$.  
When both star1 and star2 are excluded, star2 is a $2.9\sigma$ outlier, right on the boundary for inclusion/exclusion from the kinematics.  
For both stars their spatial position and metallicity  favor their membership in F6 however their velocities are either inconsistent (star1) or only marginally consistent (star2).
While we did not use the $T_{eff}$ or $\log{g}$ in our membership model we have verified that all F6 members, including star1 and star2, are consistent with being on the red-giant branch.

Due to the non-uniform spatial sampling of the spectroscopy, we explore an additional mixture model.
Instead of modeling the entirety of the Fornax dSph member, we limit our modeling to a localized sample of stars near the Fornax~6 cluster (specifically $R<10\times r_{h, {\rm F6}}$).
In this mixture model we consider the spatial distribution of the Fornax members to be approximately constant within this small region.

With this model, we find the properties of \stamp to be:
$\overline{v}_{\rm F6} = 52.1_{-2.7}^{+0.9}$,
$\sigma_{v,{\rm F6}} = 2.3_{-2.2}^{+3.9}$,
$\overline{\rm [Fe/H]}_{\rm F6} = -0.71_{-0.09}^{+0.08}$, and
$\sigma_{\rm [Fe/H], {\rm F6}} = 0.07_{-0.05}^{+0.16}$.
In this case, the velocity dispersion is not fully resolved and contains a tail in the posterior that extends to zero.
We find a total membership of $\sim10.2$ in the cluster, with 10 and 4 stars having membership larger than 0.5 and 0.9, respectively.
In this model, both star1 and star2 are considered Fornax dSph members instead of F6 members.
With both models we find $\overline{\rm [Fe/H]} \sim -0.7$ and an upper limit on the metallicity dispersion.  The velocity dispersions disagree due to the membership of the outlier stars. 
Including or excluding these stars changes the velocity dispersion and inferred mass-to-light ratio but the not the mean metallicity.
If these stars are members it is possible that they are binary stars and have inflated the velocity dispersion.
Similar cases of inflated velocity dispersion have been seen in the Bo\"{o}tes~II and Triangulum~II ultra-faint dwarf galaxies \citep{Koch2009ApJ...690..453K, Ji2016ApJ...817...41J, Kirby2015ApJ...814L...7K, Kirby2017ApJ...838...83K}. 
In both cases, small sample sizes and single epoch data led to inflated results.  Unfortunately, we only have one epoch for both star1 and star2. 
One of the main differences between the two F6 mixture models is how the Fornax spatial density is modeled; this difference results in $\sim5-7$ fewer stars considered members. This change in model/prior is effectively a different manner to account for the  spectroscopic selection function which is difficult to model and characterize.

We estimate the dynamical mass of F6 using the mass estimator: $M(<1.8 R_h)\approx 3.5 \times 1.8 R_h \sigma_{\rm los}^2 G^{-1}$ from \citet{Errani2018MNRAS.481.5073E}.
With $\sigma_{v, {\rm F6}}=5.6_{-1.6}^{+2.0}\kms$ this gives $M(<1.8 R_h)=5.2_{-2.6}^{+4.2}\times 10^{5} \Msun$, and dynamical mass-to-light ratio 
$\Upsilon (1.8 R_h) = M(1.8 r_{h})/L_V (1.8 R_h) = 96_{-48}^{+78}$ ($L_V (1.8 R_h)\approx 0.76 L_{V, ToT}$ for a Plummer profile). 
In the cases of removing the outlier stars, we find $\Upsilon (1.8 R_h) \sim 49$ and $\Upsilon (1.8 R_h) \sim 22$ when removing star1 and both stars, respectively.

Based on the F6 projected position relative to Fornax ($\Delta R \sim 270 \pc$) and its radial velocity ($\Delta v \sim 4 \kms$) it is clear that F6 is associated with the Fornax dSph.  
While the majority of its derived properties (size, luminosity, metallicity, and metallicity spread) are consistent with a globular cluster type system its velocity dispersion is larger than expected.
If F6 is a globular cluster (i.e., dark matter free), the predicted velocity dispersion is $\sigma_v\sim0.8-1.2 \kms$ for  stellar mass-to-light ratio between 1 to 2.
This is a factor of $\sim5$ less than the value we measure in our mixture model. 
$\sigma_v$ is typical for ultra-faint dwarf galaxies, however, their mass-to-light ratios tend to be larger, $\Upsilon_{1/2}>100$  \citep[e.g.,][]{Simon2019ARA&A..57..375S}. 
The other properties of F6 are not characteristic of ultra-faint dwarf galaxies.  In particular, all ultra-faint dwarf galaxies are much more metal-poor; all have  [Fe/H]$\lesssim-2$ \citep{Simon2019ARA&A..57..375S}.  
This suggests that the large velocity dispersion and mass-to-light ratio may be due to  Fornax dSph interlopers, tidal disruption, and/or unresolved binaries. 
Based on these properties we classify F6 as a globular cluster and it is the sixth globular cluster of the Fornax dSph.
This confirmation increases the globular cluster specific frequency of Fornax, which was already large for its luminosity. 
The globular cluster specific frequency is defined as $S_N = N_{GC} \times 10^{0.4 (M_V + 15)} $ \citep{Harris1981AJ.....86.1627H}. 
We find $S_N=24.8$ for Fornax (without F6 it is $S_N=20.7$).

\subsection{Permutation Tests}

\begin{figure}
\includegraphics[width=\columnwidth]{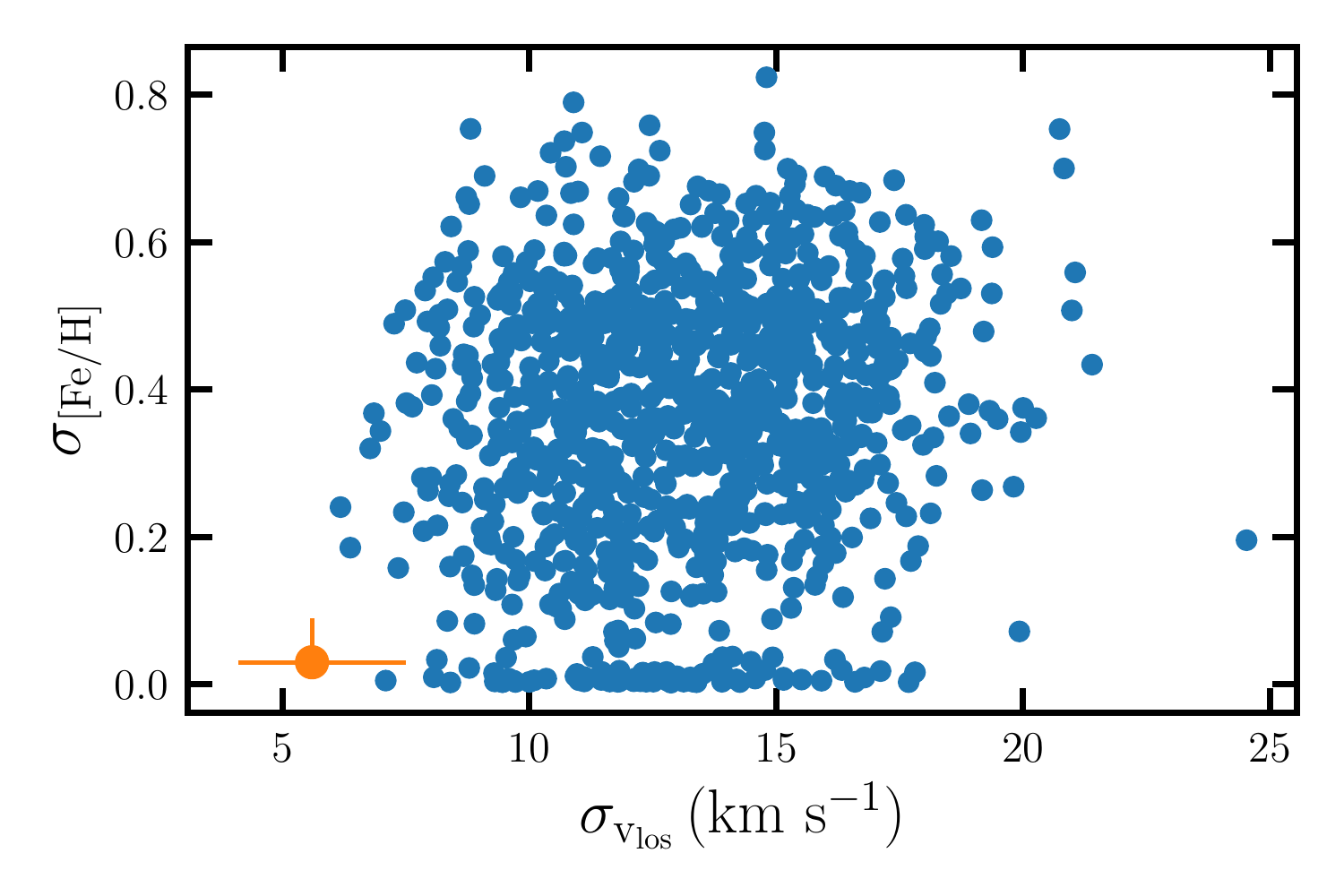}
\caption{Velocity dispersion and metallicity dispersion distribution of the 1000  permutation tests.  Each test has 12 random Fornax members with metallicity measurements and 3 additional members only with velocity measurements, mirroring the F6 sample. 
Kinematic and metallicity properties similar to a globular cluster (i.e., F6) are highly unlikely to arise from a chance alignment. 
}
\label{fig:permuation_both}
\end{figure}

To assess the significance of the detection of the F6 cluster and its properties relative to the Fornax dSph, we performed a number of data permutation tests.
We concentrate on whether we can match the low velocity and metallicity dispersions of F6 which are properties that are characteristic of globular clusters. 
For these tests we randomly drew, without replacement, a subset of stars from the Fornax dSph sample (excluding all stars within $10\times r_{h, {\rm F6}}$ of \stamp) and computed the kinematics and metallicity of this subset.
To match the properties of the F6 sample, we selected 12 stars with metallicity measurements (we additionally excluded stars with $\epsilon_{\rm [Fe/H]}>0.5$ to match the quality of the Fornax 6 members) and then 3 stars without metallicity measurements.
We additionally limited the Fornax sample to the central pointings.
We then computed the kinematic and metallicity properties of this subset and repeated this 1000 times.

In Figure~\ref{fig:permuation_both}, we examine the velocity dispersion and  metallicity dispersion  distributions from the  permutations results and compare them to the best fit F6 properties (orange point).
While there are a number of samples that are consistent within 1-sigma of the metallicity dispersion (9.1\%) or velocity dispersion (1\%) of F6, only 1 permutation (0.1\%) is consistent with both a low velocity and metallicity dispersion similar to F6. 
In summary, in the Fornax dSph, it is extremely unlikely to find by chance a clustering of stars with properties similar to F6. 
In addition, the F6 members are spatially concentrated ($R \lesssim 1\arcmin$) in contrast to the permutation tests explored.

A related question is how likely is it to find a Fornax star with F6-like velocity and metallicity properties near F6. 
The probability to find a F6-like star  drawn from a Fornax dSph-like distribution is 2-4\%.  A F6-like star is defined as $-0.79<{\rm [Fe/H]}<-0.63$ and either $46.8<v_{los}< 54.2 \kms$ or $43.2<v_{los}< 57.8 \kms$  where the difference in percentage  accounts for the potential non-zero velocity dispersion.
In the current sample there are 5 Fornax members within 3 half-light radii of F6.  If the outlier stars (star1 and star2) are considered Fornax stars and not F6 stars this would increase to 7.
If 5 stars are drawn from a Fornax-like distribution the probability that no stars have F6-like properties is 82-90\% and  for 7 stars the probability is  76-86\%. 
It is unlikely to have a Fornax star that has the F6-like properties located near the cluster but it is not excluded.  As there are two outlier F6 stars it is possible that this has occurred.

\subsection{Tidal Disruption of Fornax~6}

Depending on the line-of-sight distance to the F6 cluster relative to the Fornax dSph, there may be strong tidal forces on the cluster.  One potential signature of tidal disruption is the presence of a velocity gradient.
The disrupting MW satellite Tucana~III contains a large velocity gradient ($\sim 8 \kms {\rm deg^{-1}}$) along its short tidal tails \citep{Li2018ApJ...866...22L}.
To search for a velocity gradient we modify the mean parameter of the velocity component of the likelihood: $\overline{v} \rightarrow \overline{v} + v_g \chi(\theta_g)$. Here, $v_g$ is the velocity gradient magnitude and $\chi(\theta_g)$ is the distance along the velocity gradient axis which is rotated by an angle $\theta_g$.
This adds two additional velocity parameters ($v_g$, $\theta_g$).
We  apply this model to stars identified as F6 members and weight each star by its membership probability.
For simplicity we only  model the velocity distribution.
We use the MCMC code \texttt{emcee} to sample the posterior \citep{ForemanMackey2013PASP..125..306F}. 

With the F6 stars identified above we find  a gradient that is consistent with zero ($v_g=2.7\pm4.4 \kms {\rm arcmin^{-1}}$).  If we remove the two outlier stars the significance increases ($v_g=2.6\pm2.7 \kms {\rm arcmin^{-1}}$) but is still consistent with no gradient. 
If we use the members identified with the second F6 mixture model, we find $v_g=4.5_{-2.2}^{+2.1} \kms {\rm arcmin^{-1}}$ with $\theta_g=-36\pm20\degree$. 
We do not find any strong evidence for the presence of a velocity gradient.
If there is a velocity gradient a larger spectroscopic sample is required to identify it.

To estimate the tidal force of  Fornax  on F6 we compute the Jacobi radius, $r_j = (M_{\rm F6}/M_{dSph}(r_{\rm F6})/3)^{1/3} r_{\rm F6}$ \citep{BT2008}.
Assuming an NFW halo for  Fornax  with a scale radius of 3 kpc and a scale density of  $0.044 \, \Msun {\rm pc^{-3}}$, we find $r_j\sim17-29~\pc$ if F6 is a purely stellar system with $\Upsilon=2$, and $r_j\sim60-100~\pc$ if instead the measured velocity dispersion is representative of the mass profile of F6 ($\Upsilon \sim 100$).  In both cases, the range denotes a line-of-sight distance between 0-1 kpc and a projected distance of 270 pc. 
For a purely stellar F6, the Jacobi radius is smaller than the size of the system, suggesting that F6 should be undergoing tidal disruption, which may explain its larger elongation compared to the other Fornax GCs. 
If the larger velocity dispersion is representative of F6, than the cluster should be bound as the Jacobi radius is larger than the cluster. 
Any conclusions drawn about the Jacobi radii and potential tidal disruption depend on the membership of the two outlier candidates (star1 and star2). 
We note that $r_J$ will be larger if Fornax has a cored profile \citep{Wang2019ApJ...875L..13W}.
If the true velocity dispersion is on the low end or there is some inflation of the velocity dispersion due to a velocity gradient then F6 could be disrupting but if the high velocity dispersion is representative then the cluster members would be bound.

It has been found that in the Fornax system, a large fraction of the metal-poor stars belong to the 4 metal-poor globular clusters, up to 20-25\% \citep{Larsen2012A&A...544L..14L}.
It is therefore interesting to determine the fraction of the metal-rich stars ([Fe/H]$>-1$) located in F6 compared to the number of metal-rich stars in the Fornax dSph. 
To account for the sampling bias we compare the stellar mass in F6 to the fraction of  stellar mass in Fornax that is metal-rich. In our M2FS sample, we find 57\% of the Fornax stars have [Fe/H]$>-1$. Using  the $M_V$ values in Table~\ref{tab:for6_properties} and assuming that both F6 and Fornax have the same stellar M/L, we find that $\sim0.6\%$ of the metal-rich stars in the Fornax system are located within the F6 cluster.  
This estimation may  be biased if the  M2FS metallicity distribution is not  representative of the Fornax  metallicity distribution due to spatial sampling.
In summary, we find that just $\sim0.6\%$ of the metal-rich stars belong to the F6 cluster, which  is a much lower than the fraction of metal-poor stars that belong to the metal-poor globular clusters.  

One possible reason for the difference in globular cluster stellar fraction between F6 and the metal-poor clusters is that more F6 stars have been tidally stripped.  We have examined the spatial distribution of stars with metallicity similar to F6 ($-0.9 < {\rm [Fe/H]}<-0.5$) in the M2FS sample to search for direct evidence of mass loss.  We see no obvious concentration of metal-rich stars around the F6 cluster nor tidal tail like features originating from the cluster.  
The F6-like stars are more centrally concentrated than the bulk of the Fornax sample but this subset of stars follows the general trend of observing more metal-rich stars in the central regions of Fornax (i.e., a negative metallicity gradient).
If there are disrupted F6 stars in the bulk of Fornax, a larger sample of stars with more precise metallicities  or light element abundance measurements are required to identify them.  For example, a tidally stripped M54 star has been identified in the Sagittarius dSph based on its nitrogen and aluminum abundances  \citep{FernandezTrincado2021A&A...648A..70F}.

\subsection{Age}

\begin{figure}
\includegraphics[width=\columnwidth]{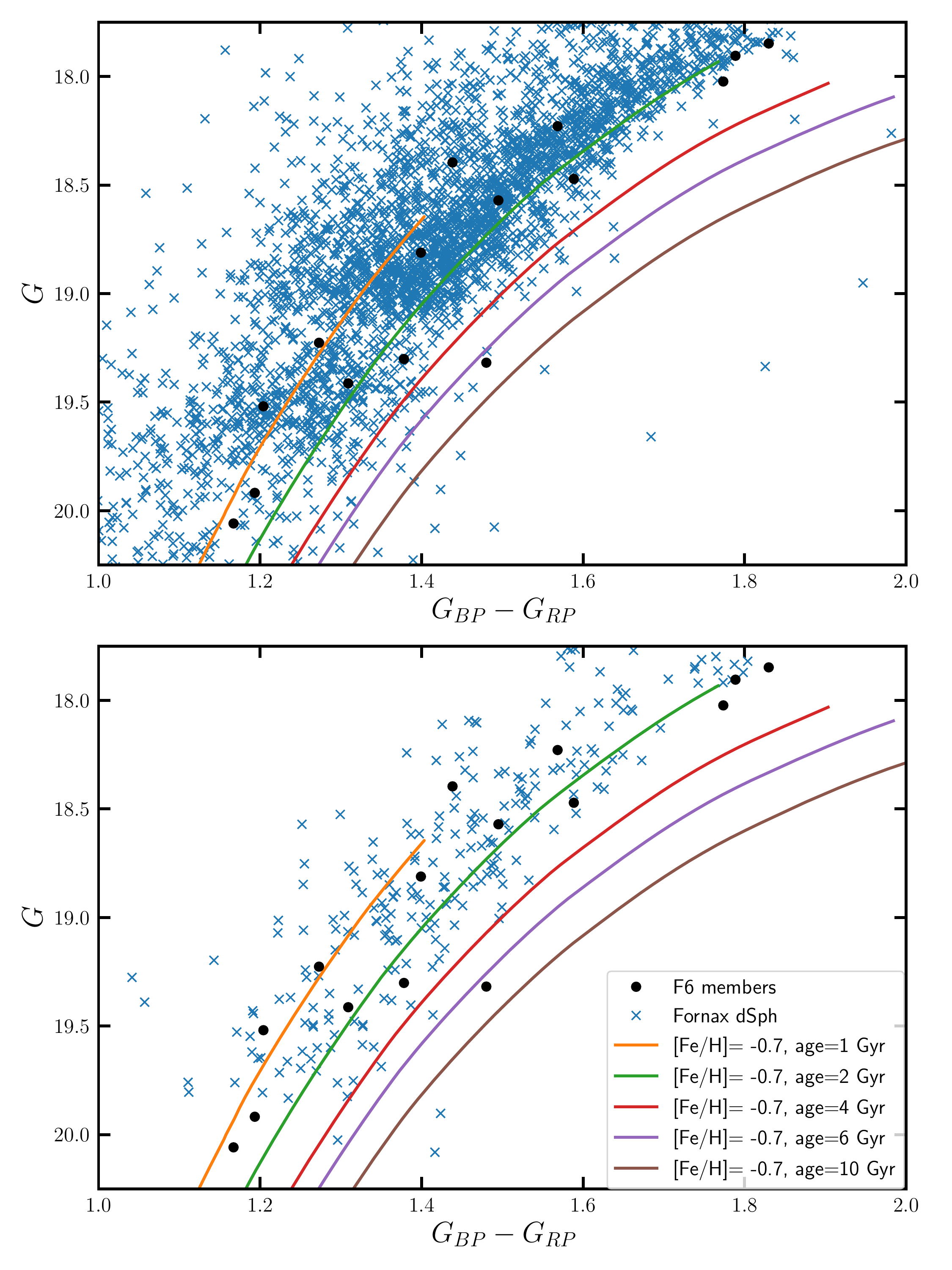}
\caption{Color-magnitude diagram of F6 members (black points) and  Fornax dSph members (blue x).  The top panel shows all Fornax dSph members and the lower panel shows only Fornax dSph members within $R_{F6}<5\arcmin$. Overlaid are MIST isochrones with a fixed metallicity based on the F6 members ([Fe/H]=-0.7) and ages at 1, 2, 4, 6, and 10 Gyr. 
}
\label{fig:cmd_isochrone}
\end{figure}

One of the last remaining pieces of the F6 properties is the age of the system.
In Figure~\ref{fig:cmd_isochrone}, we compare the colors and magnitudes of F6 members in $G_{BP}-G_{RP}$ vs $G$  to several theoretical isochrones from the MESA Isochrones and Stellar Tracks  \citep[MIST;][]{Dotter2016ApJS..222....8D, Choi2016ApJ...823..102C}. 
All isochrones are fixed to [Fe/H]=-0.7, the best-fit metallicity measured from the F6 members, and we vary the age of the isochrone (1, 2, 4, 6, and 10 Gyr). For comparison the top panel includes all Fornax members and the bottom panel includes nearby Fornax members within $R_{\rm F6}<5\arcmin$. We note that the Fornax dSph members span a range in both metallicity and age.
The isochrone with an age of 2 Gyr provides the best approximate fit for the F6 stars. If we use $G-G_{RP}$ instead of $G_{BP}-G_{RP}$ we reach a similar conclusion. 

Work on the star formation history of the Fornax dSph measured with  deep HST photometry found a major  star formation episode at old ages ($>10$ Gyr), with a second recent burst at  $\sim4.6$ Gyr, and intermittent star formation in recent times $\sim 0.2-2$ Gyr \citep{Rusakov2021MNRAS.502..642R}.
The age estimation based on the comparison to the isochrones matches the more recent intermittent star formation.  In addition, the photometric metallicity from the star formation histories of  \citet{Rusakov2021MNRAS.502..642R} roughly matches $[Fe/H]\sim -0.7$ at 2 Gyr.
The Fornax dSph star formation episode at $\sim2$ Gyr may have  triggered the conditions necessary to form the F6 cluster. 
The most recent pericenter of Fornax based on {\it Gaia} DR2 proper motions is $t_p=1.7_{-0.3}^{+1.7}$ Gyr which corresponds to the most recent star formation \citep{Rusakov2021MNRAS.502..642R}.
The previous pericenter time corresponds roughly to the $4-6$ Gyr burst.

There is evidence of past mergers in the kinematics and chemistry of the Fornax stars.
For example, the red-giant branch split is split into 2-3 chemo-dynamic populations \citep{Battaglia2006A&A...459..423B, Walker2011ApJ...742...20W, Amorisco2012ApJ...756L...2A}.
These three groups have $\overline{\rm [Fe/H]}\sim -1.8, -1, -0.65$. which may correspond to the inital old star formation episode ($>10$ Gyr), the second burst ($\sim4.6$ Gyr), and the most recent burst ($\sim 0.2-2$ Gyr) \citep{Rusakov2021MNRAS.502..642R}. 
\citet{Amorisco2012ApJ...756L...2A} found that the intermediate population and the metal-rich population have a misalignment between their angular momentum vectors suggesting that Fornax is a merger of a bound pair.
\citet{delPino2017MNRAS.465.3708D} find rotating sub-populations in the Fornax chemo-dynamics and a change in direction for stars of $\sim 8$ Gyr suggests that a dwarf galaxy merger occurred around this time.
Similar evidence is found by the misalignment between the youngest stars in Fornax and the old stellar population \citep{Wang2019ApJ...881..118W}.

The history of the Fornax dSph is complex, with multiple episodes of star formation, some evidence of mergers, and pericenter passages that may correspond to the formation time of F6. 
While we have provided a simple estimate of the age, future more accurate age measurements are required to conclusively identify any of these events with F6.

\subsection{Comparison to Other Globular Clusters}

\begin{figure*}
\includegraphics[width=\textwidth]{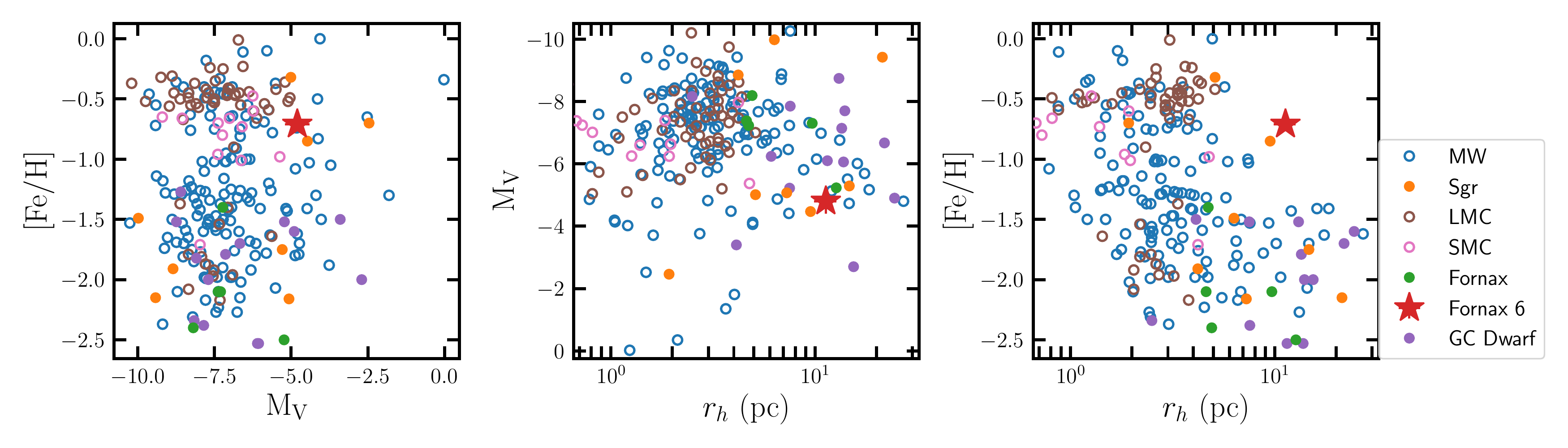}
\caption{Comparison of the F6 cluster (red) to globular clusters of the MW (blue), Sgr dsph (orange), Fornax dSph (green), {Large Magellanic Cloud (LMC; brown), Small Magellanic Cloud (SMC; pink)} and GC of other dwarf galaxies in the local group (purple; Eridanus II, And I, And XXV, WLM, Sextans A, Pegasus dIrr, and NGC 6822).
From left to right the three panels show: absolute magnitude ($M_V$) versus metallicity ([Fe/H]), half-light radius  ($r_{h}$) versus absolute magnitude ($M_V$), and half-light radius  ($r_{h}$)  versus metallicity ([Fe/H]).
}
\label{fig:gc_compare}
\end{figure*}

\begin{figure}
\includegraphics[width=\columnwidth]{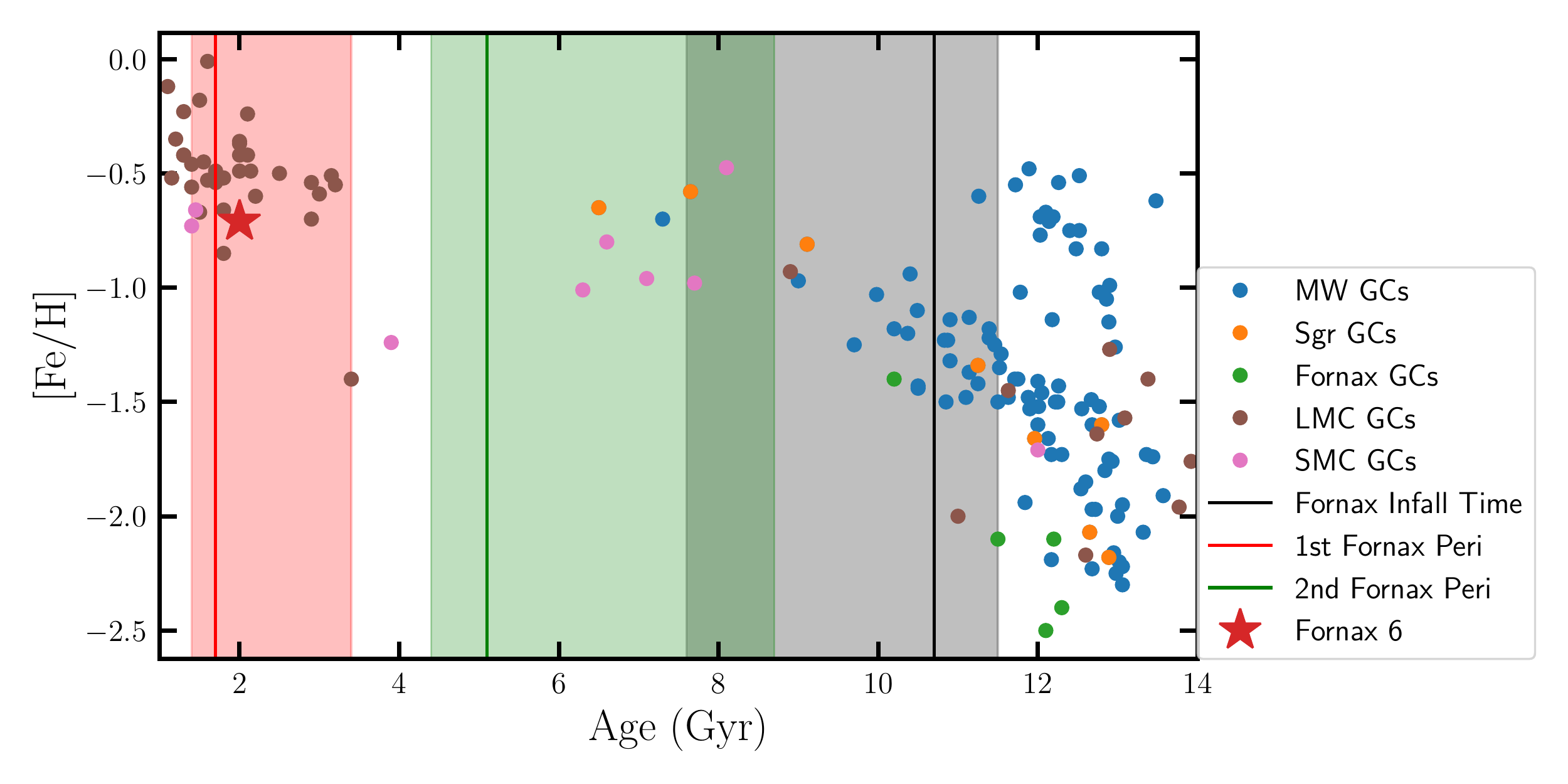}
\caption{Age-metallicity distribution for Galactic, {LMC, SMC,} Sagittarius, and Fornax star clusters.  
The old- metal-rich clusters are the in-situ MW clusters and the track of old-metal-poor to intermediate age and higher metallicities are accreted globular clusters.  The Fornax and Sagittarius globular clusters follow age-metallicity trend of  accreted globular clusters.
{The LMC and SMC host a sizable population of clusters with similar age and metallicity to F6.}
The red and green shaded bands correspond to the most recent and second most recenter pericenter passage of Fornax \citep{Rusakov2021MNRAS.502..642R}.
The black shaded band shows the infall time of the Fornax dSph into the MW,  $t_{\rm infall} = 10.7_{-3.1}^{+0.8}$ \citep{Fillingham2019arXiv190604180F}.
}
\label{fig:age_feh}.
\end{figure}

The main peculiarity  of F6 compared to other Fornax GCs is that it is much more metal-rich than the other five Fornax clusters. 
In Figure~\ref{fig:gc_compare}, we compare the luminosity, metallicity, and half-light radius of F6 to other GCs in the Local Group. 
This includes:  the other five Fornax GCs \citep{Mackey2003MNRAS.340..175M, Letarte2006A&A...453..547L, Larsen2012A&A...546A..53L}, the Sagittarius (Sgr) dSph\footnote{The Sgr dwarf is undergoing strong tidal disruption from the MW and has prominent tidal tails \citep[e.g.,][]{Majewski2003ApJ...599.1082M, Ramos2020A&A...638A.104R}.
There are four Sgr GCs associated with the core of the galaxy (Arp~2, NGC~6715, Terzan~7, Terzan~8) \citep{DaCosta1995AJ....109.2533D} and four  associated with the tidal tails (NGC~2419, NGC~5824, Palomar~12, and Whiting~1) \citep[e.g.,][]{Law2010ApJ...718.1128L, Massari2019A&A...630L...4M, Kruijssen2020MNRAS.498.2472K}. 
There are other GCs suggested to be members of Sgr system associated with the tidal tails but the membership of these GCs is more tentative (NGC 4147, Pal 2, NGC 6284) \citep[e.g.,][]{Law2010ApJ...718.1128L, Bellazzini2020A&A...636A.107B}.} (Arp~2, NGC~2419, NGC~5824, NGC~6715, Palomar~12, Terzan~7, Terzan~8, and Whiting~1), the MW \citep[][2010 edition]{Harris1996AJ....112.1487H}, the Large Magellanic Cloud \citep[LMC;][]{Mackey2003MNRAS.338...85M, Pessev2006AJ....132..781P, Pessev2008MNRAS.385.1535P, Song2021MNRAS.504.4160S}, Small Magellanic Cloud \citep[SMC;][]{Mackey2003MNRAS.338..120M, Pessev2006AJ....132..781P, Pessev2008MNRAS.385.1535P, Song2021MNRAS.504.4160S} , and other dwarf galaxy hosts 
(WLM \citep{Hodge1999ApJ...521..577H, Stephens2006AJ....131.1426S}, Eridanus~II \citep{Simon2021ApJ...908...18S}, And~I \citep{Caldwell2017PASA...34...39C}, And XXV \citep{Cusano2016ApJ...829...26C}, Sextans~A \citep{Beasley2019MNRAS.487.1986B}, Pegasus dIrr \citep{Cole2017ApJ...837...54C, Leaman2020MNRAS.492.5102L}, and NGC~6822 \citep{Hwang2011ApJ...738...58H, Hwang2014ApJ...783...49H}).
F6 has a large size compared to other GCs at a similar metallicity 
Almost all GCs with [Fe/H]$>-1.25$ have small sizes ($r_h < 5 \pc$).
The GC in the ultra-faint dwarf galaxy Eridanus~II is analogous in size and luminosity as F6 but the Eri~II GC metallicity is more metal-poor \citep[${\rm [Fe/H]}\approx -2$;][]{Zoutendijk2020A&A...635A.107Z, Simon2021ApJ...908...18S}.
The closest direct analog to F6 is the Palomar~12 GC, which is a Sgr GC located in the trailing arm of the Sgr stream. 
The Sgr dSph has three more metal-rich clusters, Palomar~12, Whiting~1, and Terzan~7, similar to F6. 
The LMC and SMC both have younger and more metal-rich clusters compared to the dwarf galaxy hosts and their sizes are  similar to MW GC population.
The large size of F6 relative to the LMC/SMC clusters could be due to the recent formation and smaller tidal field from the Fornax dSph versus  the massive LMC.  
Excluding the Sgr GCs, the other dwarf galaxy GCs are  generally more metal-poor than the MW distribution in contrast to F6.

In Figure~\ref{fig:age_feh}, we show the age-metallicity distribution of  globular clusters in the Fornax dSph, Sagittarius dSph, LMC, SMC, and MW \citep[age measurements are from][]{deBoer2016A&A...590A..35D, Kruijssen2019MNRAS.486.3180K, Song2021MNRAS.504.4160S}.
We include shaded bands showing the past two pericenter passages \citep{Rusakov2021MNRAS.502..642R} and the infall time of the Fornax dSph into the MW, $t_{\rm infall} = 10.7_{-3.1}^{+0.8}$ \citep{Fillingham2019arXiv190604180F}. Both pericenter passages correspond to increased star formation in the Fornax dSph \citep{Rusakov2021MNRAS.502..642R}.
The MW GCs form two tracks, an old- metal-rich clump (in-situ) (upper right) and an accreted track where the GCs are correlated in their age and metallicity.
If we assume F6 follows the same age-metallicity  accreted relation as the MW, we would expect an age similar to the young Sgr GCs,  in the range of  $\sim6-10\Gyr$.
In particular, Palomar~12 has an age $\sim9\Gyr$ \citep{Kruijssen2019MNRAS.486.3180K}.  We note that all three metal-rich Sgr GCs have ages $\sim9$ Gyr or younger. 
This age range overlaps with the infall time of the Fornax dSph \citep{Fillingham2019arXiv190604180F}, and suggests there could be a relation between the formation of F6 and the infall of the dSph.
However, there is better agreement between the age estimated from the {\it Gaia} color-magnitude diagram and  the most recent pericenter passage of Fornax.
Ultimately, accurate photometry of the sub-giant and main-sequence turn-off is required for a more accurate age measurement of F6 and to determine whether the formation of F6 is related to the orbital history of the Fornax dSph.

\section{Conclusion}
\label{label:conclusion}

We present Magellan/M2FS spectroscopy of the Fornax dSph and the Fornax~6 (F6) globular cluster.
We measure 980 new radial velocities and metallicities of stars which we combined with previous MMFS spectroscopy \citep{Walker2009AJ....137.3100W} and {\it Gaia} EDR3 astrometry.
With this spectroscopic sample we are able to identify members of the F6 cluster and Fornax dSph.
Our main findings are as follows:

\begin{itemize}
    \item We identify 2989 members of the Fornax dSph and measure a velocity dispersion of $12.1\kms$, a mean metallicity of $-1.22$, a metallicity dispersion of 0.48 and a metallicity gradient of $-0.28 \, {\rm dex \, R_{h}^{-1}}$.   
With an efficient {\it Gaia} DR2 target selection, a number of these stars (165) are at large radii ($R/r_h > 2$). This is one of the largest Fornax samples and will be useful for future kinematic analyses. 

\item We identify 15-17 members of the F6 cluster and confirm that F6 is distinct in velocity and metallicity from Fornax dSph stars (see Figure~\ref{fig:for6_diagnostic}).
We measure $\overline{v}_{\rm F6} = 50.5_{-1.7}^{+1.7} \kms$ and $\sigma_{v, {\rm F6}} = 5.6_{-1.6}^{+2.0} \kms$.  
The small offset in radial velocity between F6 and the Fornax dSph combined with its spatial location confirms that the F6 is associated with the Fornax dSph. 
There are two velocity outlier stars that may be inflating $\sigma_{v, {\rm F6}}$. If they are excluded from the Fornax~6 sample, the velocity dispersion is consistent with zero.
We measure $\overline{\rm [Fe/H]}_{\rm F6} = -0.71_{-0.05}^{+0.05}$ and $\sigma_{\rm [Fe/H], {\rm F6}} = 0.03_{-0.02}^{+0.06}$ which is more metal-rich than the bulk of the Fornax dSph stars and the other five Fornax GCs. 
We estimate that 0.6\% of the metal-rich stars are located within the F6 cluster whereas a much larger percentage of the metal-poor stars in the Fornax system are located within the other GCs \citep[20-25\% are located within F1,F2,F3,F5;][]{Larsen2012A&A...544L..14L}.
Excluding the velocity dispersion,  the metallicity, size ($r_{h, {\rm F6}}\sim11.3\pc$), and luminosity ($M_V\sim-4.8$) of F6 suggest that it is a GC. 

\item We have compared F6 to other GCs in the MW and other dwarf galaxy systems (see Figure~\ref{fig:gc_compare}). 
When compared to other metal-rich GCs, F6 has the largest physical size.  However, its size and luminosity is similar to other metal-poor GCs in the MW and dwarf galaxy GCs.
F6 has properties similar to the metal-rich GCs of Sagittarius.  In particular, Palomar~12 is an excellent analog based on [Fe/H], $R_h$, and $M_V$. 

\item 
Given the metal-rich nature of the F6 cluster ([Fe/H]$\sim-0.7$), it is almost certainly younger than the other Fornax GCs (age$<10\Gyr$) which places the formation after infall into the MW \citep[$t_{if}\sim 10.7 \Gyr$;][]{Fillingham2019arXiv190604180F}.
From {\it Gaia} photometry and MIST isochrones we estimated an age of 2 Gyr.
Fornax has an extended star formation history \citep{Rusakov2021MNRAS.502..642R} and potential dwarf galaxy mergers \citep{Amorisco2012ApJ...756L...2A, delPino2017MNRAS.465.3708D}.  
There are star bursts at $\sim 2,4 \Gyr$ that correspond to the two most recent pericenter passages \citep{Rusakov2021MNRAS.502..642R, Fritz2018A&A...619A.103F}.  
Any of these events could have triggered the necessary conditions for the formation of the F6 cluster.  
More precise photometry is required to improve the measurement of the age and identify the origin of the cluster.
\end{itemize}

The survival of the Fornax globular cluster system is arguably at odds with dynamical fictions arguments and has been used as evidence for a cored dark matter halo of the Fornax dSph.
Future analysis of this problem should include F6, as the inner clusters (F3, F4, F6) are the most impacted by different dark matter halos. 
While the systemic proper motion of F6 relative to Fornax is currently not well constrained, in future {\it Gaia} data releases the proper motions will be more informative and assist with addressing the `Fornax timing problem.'

\acknowledgments
ABP is supported by NSF grant AST-1813881.  M.G.W. acknowledges support from NSF grants AST-1813881 and AST-1909584. SK  was partially supported by NSF grants AST-1813881 and AST-1909584.
EO was partially supported by NSF grant AST-1815767.
NC is supported by NSF grant AST-1812461.
MM was supported by U.S.\ National Science Foundation (NSF) grants AST-1312997, AST-1726457 and AST-1815403.  
We thank the referee for their helpful comments. 

This work has made use of data from the European Space Agency (ESA)
mission {\it Gaia} (\url{https://www.cosmos.esa.int/gaia}), processed by the {\it Gaia} Data Processing and Analysis Consortium (DPAC,
\url{https://www.cosmos.esa.int/web/gaia/dpac/consortium}). Funding
for the DPAC has been provided by national institutions, in particular
the institutions participating in the {\it Gaia} Multilateral Agreement.

This research has made use of NASA's Astrophysics Data System Bibliographic Services.
This paper made use of the Whole Sky Database (wsdb) created by Sergey Koposov and maintained at the Institute of Astronomy, Cambridge by Sergey Koposov, Vasily Belokurov and Wyn Evans with financial support from the Science \& Technology Facilities Council (STFC) and the European Research Council (ERC).

This project used public archival data from the Dark Energy Survey (DES). Funding for the DES Projects has been provided by the U.S. Department of Energy, the U.S. National Science Foundation, the Ministry of Science and Education of Spain, the Science and Technology FacilitiesCouncil of the United Kingdom, the Higher Education Funding Council for England, the National Center for Supercomputing Applications at the University of Illinois at Urbana-Champaign, the Kavli Institute of Cosmological Physics at the University of Chicago, the Center for Cosmology and Astro-Particle Physics at the Ohio State University, the Mitchell Institute for Fundamental Physics and Astronomy at Texas A\&M University, Financiadora de Estudos e Projetos, Funda{\c c}{\~a}o Carlos Chagas Filho de Amparo {\`a} Pesquisa do Estado do Rio de Janeiro, Conselho Nacional de Desenvolvimento Cient{\'i}fico e Tecnol{\'o}gico and the Minist{\'e}rio da Ci{\^e}ncia, Tecnologia e Inova{\c c}{\~a}o, the Deutsche Forschungsgemeinschaft, and the Collaborating Institutions in the Dark Energy Survey.
The Collaborating Institutions are Argonne National Laboratory, the University of California at Santa Cruz, the University of Cambridge, Centro de Investigaciones Energ{\'e}ticas, Medioambientales y Tecnol{\'o}gicas-Madrid, the University of Chicago, University College London, the DES-Brazil Consortium, the University of Edinburgh, the Eidgen{\"o}ssische Technische Hochschule (ETH) Z{\"u}rich,  Fermi National Accelerator Laboratory, the University of Illinois at Urbana-Champaign, the Institut de Ci{\`e}ncies de l'Espai (IEEC/CSIC), the Institut de F{\'i}sica d'Altes Energies, Lawrence Berkeley National Laboratory, the Ludwig-Maximilians Universit{\"a}t M{\"u}nchen and the associated Excellence Cluster Universe, the University of Michigan, the National Optical Astronomy Observatory, the University of Nottingham, The Ohio State University, the OzDES Membership Consortium, the University of Pennsylvania, the University of Portsmouth, SLAC National Accelerator Laboratory, Stanford University, the University of Sussex, and Texas A\&M University.
Based in part on observations at Cerro Tololo Inter-American Observatory, National Optical Astronomy Observatory, which is operated by the Association of Universities for Research in Astronomy (AURA) under a cooperative agreement with the National Science Foundation.

\vspace{5mm}
\facilities{Magellan}

\software{astropy \citep{Astropy2013A&A...558A..33A, Astropy2018AJ....156..123A},
\texttt{matplotlib} \citep{matplotlib}, 
\texttt{NumPy} \citep{numpy},
\texttt{iPython} \citep{ipython},
\texttt{SciPy} \citep{scipy}
\texttt{corner.py} \citep{corner}, 
\texttt{emcee} \citep{ForemanMackey2013PASP..125..306F}  ,
\texttt{Q3C} \citep{2006ASPC..351..735K}
          }

\bibliography{main_bib_file}{}
\bibliographystyle{aasjournal}

\end{document}